\documentclass[a4paper,11pt]{article}
\immediate\write18{rm -f *_cache.pdf}
\usepackage{jheppub} 
\usepackage{lineno}

\usepackage{amsmath}
\usepackage{amsfonts}
\usepackage{gensymb}
\usepackage{amssymb}
\usepackage{graphicx}
\usepackage{pdfpages}
\usepackage{multirow}
\usepackage{empheq}
\usepackage{subcaption}
\usepackage{dirtytalk}

\title{\boldmath Simultaneous production of a $W$ boson and a charmed hadron at the LHC in general-mass variable-flavour-number scheme}

\author{Ville Alanko, Ilkka Helenius, Hannu Paukkunen}
\affiliation{Department of Physics, University of Jyväskylä,
P.O. Box 35, FI-40014 University of Jyväskylä, Finland\\
Helsinki Institute of Physics, 
P.O. Box 64, FI-00014 University of Helsinki, Finland}

\emailAdd{alankovh@jyu.fi, ilkka.m.helenius@jyu.fi, hannu.t.paukkunen@jyu.fi}

\abstract{The simultaneous production of a $W^\pm$ boson and a charmed hadron
in proton-proton collisions
offers a potential probe for constraining the strange quark parton distribution functions (PDFs). We study these processes at next-to-leading order in perturbative Quantum Chromodynamics within the general-mass variable-flavor-number scheme. By considering ratios of cross-section between oppositely charged mesons, uncertainties associated with unphysical scale choices and fragmentation functions are shown to effectively cancel out, leaving the uncertainty originating from the PDFs as the dominant one. 
By comparing our calculations with CT18A, MSHT20 and NNPDF4.0 PDFs with the recent ATLAS measurement at $\sqrt{s} = 13\,{\rm TeV}$ we find that CT18A, which imposes zero strangeness asymmetry, agrees best with the data, while MSHT20 and NNPDF4.0, both of which allow for a non-zero strangeness asymmetry, exhibit greater tension with the ATLAS data. The sensitivity to the strangeness asymmetry is further confirmed by the PDF reweighting methods. We also study the impact of possible intrinsic charm content of the proton finding no significant sensitivity.
  Finally, we explore the possibility of measuring these processes in proton-lead collisions. With the estimated detector efficiencies and projected luminositites at the high-luminosity LHC, these processes should be visible, yet with a rather limited constraining power for nuclear PDFs.
 }


\begin{document}
\maketitle
\flushbottom

\section{Introduction}
In high-energy particle colliders, such as the Large Hadron Collider (LHC), hadrons are commonly encountered either as the particles being collided or as the particles being produced and measured directly or indirectly. In the parton model, hadrons consist of elementary particles, partons, which in Quantum Chromodynamics (QCD) are quarks and gluons. Parton distribution functions (PDFs) \cite{Kovarik:2019xvh,Ethier:2020way,Klasen:2023uqj} quantify the number distributions of partons inside hadrons, typically protons and other nuclei, as a function of the fractional momentum $x$ they carry. Similarly, fragmentation functions (FFs) \cite{Albino:2008gy,Metz:2016swz} describe the distributions of hadrons as decay products of partons. The cross section for a process involving hadrons in the initial and final state can be calculated as a convolution between PDFs, the partonic cross section, and
FFs. The assumption that the cross section organizes in this way into parts encoding the long-distance physics (PDFs and FFs) and the short-distance physics (partonic cross section)
is called \textit{collinear factorization} \cite{Collins:1989gx,Sterman:2022gyf}. Although it has only been proven for certain processes, it is conjectured to apply more broadly -- something that is
supported by agreement with experimental data in the case of many processes \cite{Chiefa:2025loi}.

While the partonic cross sections can be calculated by perturbative methods in QCD as power series in the strong coupling, PDFs and FFs are non-perturbative in nature and it is currently not possible to derive them precisely
from first principles \cite{Lin:2017snn,Constantinou:2020hdm}. Instead, they are extracted through global analysis of collision processes by finding the form of the distributions such that the calculations optimally match the data. In this analysis, PDFs and FFs are assumed to be \textit{universal}, meaning that they are process independent and depend only on the species of the parton and the hadron, the momentum fraction of the parton and the energy scale separating the long- and short-distance physics. The dependence of PDFs and FFs as a function of this factorization scale $\mu_{\rm fact}$ is predicted succesfully by the \textit{Dokshitzer-Gribov-Lipatov-Altarelli-Parisi} (DGLAP) equations \cite{DGLAP1, DGLAP2, DGLAP3, DGLAP4}.

A significant effort has been put into the determination of
the proton PDFs \cite{Kovarik:2019xvh,Ethier:2020way}. Traditionally, a more difficult component to constrain has been the strange quark PDF which typically carries larger uncertainties than the other quark PDFs \cite{Faura:2020oom}.
An open question is the size of \textit{strangeness asymmetry}, which refers to $s-\Bar{s}$ PDF not being zero for all values of the momentum fraction $x$ while still yielding zero when integrated over $x$. The next-to-next-to-leading order (NNLO) DGLAP evolution is known \cite{Catani_2004} to create a small perturbative asymmetry but no conclusive evidence of a significant non-zero strangeness asymmetry at the PDF parametrization scale $\mu_{\rm fact} \sim 1\,{\rm GeV}$ exist to date \cite{Hou:2022onq,Anderson:2024evk} and some PDF analyses such as CT18 \cite{Hou:2019efy} set the asymmetry directly to zero, while others such as MSHT20 \cite{MSHT20PDF} and NNPDF4.0 \cite{NNPDF40PDF} let it vary freely. 
Another open question is the existence of \textit{intrinsic charm quarks} \cite{Brodsky_2015,Hou:2017khm,Ball:2022qks}. In the language of PDFs, this would mean that the charm-quark PDFs would not only be generated perturbatively through $g \rightarrow c \Bar{c}$ splittings. Intrinsic charm could manifest another asymmetry, $c\neq \Bar{c}$ \cite{NNPDF:2023tyk}.

 Processes that carry sensitivity to the the strange-quark distributions include semi-inclusive kaon production in deep-inelastic scattering (DIS) of electrons and protons \cite{Airapetian_2008,HERMES:2013ztj,Borsa:2017vwy,Sato:2019yez}, charm production in charged-current electron-proton DIS \cite{ZEUS:2019oro}, dimuon production in charged-current neut\-rino-nucleus scattering \cite{charged_current_neutrino_1, charged_current_neutrino_2, charged_current_neutrino_3, charged_current_neutrino_4, charged_current_neutrino_5} and inclusive vector-boson measurements in proton-proton ($pp$) collisions \cite{Kusina:2012vh,ATLAS_data_s_PDF_1, ATLAS_data_s_PDF_2,ATLAS:2012sjl}. The interpretation of these data have not been unique, but e.g. an optimal match with the LHC vector-boson measurements has traditionally required a larger strange-quark PDF than what the dimuon data prefer \cite{Faura:2020oom}. The $W^\pm + {\rm charm}$ production in $pp$ collisions offers a complementary process to study the strange-quark content. The idea of using $W^\pm \, + $ charm production in hadronic collisions to constrain the strange-quark content of nucleons was originally proposed in Ref.~\cite{Baur:1993zd} leading to the first experimental studies at Tevatron \cite{CDF:2007raw,D0:2008ygk,CDF:2012mhm} and to more recent studies at the LHC \cite{CMS:2013wql,LHCb:2015bwt,ATLAS:2014jkm,CMS:2018dxg,measurementforcomparison}. Out of the global fits of proton PDFs, e.g. the NNPDF4.0 \cite{NNPDF40PDF} and MSHT20 \cite{MSHT20PDF} analyses already include some $W^\pm + {\rm charm \ jet}$ LHC data from ATLAS \cite{ATLAS:2014jkm} and CMS \cite{CMS:2013wql,CMS:2018dxg} in varying combinations, and also CT18 performs consistency checks against such data \cite{Hou:2019efy}.

In this article, we study a family of processes in which a $W^\pm$ boson, subsequently decaying into a charged lepton $\ell^\pm$ and its (anti)neutrino $\overline{\nu}_l/\nu_l$, and a charmed hadron are produced simultaneously in $pp$ collisions. In particular, we will consider
\begin{equation}
    \label{eq: p p > W D}
    p + p \rightarrow W^\mp (\rightarrow l ^\mp + \Bar{\nu}_l / \nu_l) + D^{(*)\pm} + X,
\end{equation}
where $D^{(*)\pm}$ is a shorthand for $D^{\pm}$ mesons and their excited states $D^{*\pm}(2010)$. 
The processes \eqref{eq: p p > W D} can be approached from different theoretical perspectives. The simplest option is to perform a differential parton-level calculation $pp \rightarrow Wc + X$ retaining the full quark mass dependence \cite{Giele:1995kr} and handle the hadronization by folding the result with a scale-independent charm-to-hadron FF. However, at next-to-leading order (NLO) and beyond such calculation will contain large logarithmic contributions $\sim \log(p_{T, c}/m_c)$ \cite{Giele:1995kr} where $p_{T, c}$ refers to the transverse momentum of the charm quark and $m_c$ to the charm-quark mass, which originate from collinear radiation of gluons from final-state charm quarks, and the result is reliable only at sufficiently low values of $p_{\rm T,c}$. To obtain a stable result these logarithmic contributions need to be resummed. This can be achieved either by matching the quark-level calculations with a parton-shower algorithm, typically handled through an event-generator like Pythia \cite{Bierlich:2022pfr} which also takes care of hadronizing the quarks \cite{Bevilacqua:2021ovq}. Another option to sum the logarithmic terms is to introduce scale-dependent parton-to-hadron FFs \cite{KKKS08,Anderle:2017cgl,Soleymaninia:2017xhc,Salajegheh:2019nea}. This latter approach falls under the umbrella of general-mass variable-flavour-number scheme (GM-VFNS) \cite{Olness:1987ep,Aivazis:1993kh,Aivazis:1993pi,Kretzer:1997pd,Collins:1998rz,Kramer:2000hn,Tung:2001mv,Kretzer:2003it,Thorne:2008xf,Guzzi:2011ew,Helenius:2018uul,Gao:2021fle,Guzzi:2024can,Risse:2025smp,Helenius:2025fpy,Helenius:2026uuz}, and it is this approach we are going to adopt in the present paper at the NLO accuracy. At sufficiently high values of transverse momentum the GM-VFNS results reduce asymptotically to the zero-mass calculations which have very recently been calculated up to next-to-next-to-leading order (NNLO) in perturbative QCD \cite{Caletti:2024xaw,NNLOcalculation}.

In what follows, in Section~\ref{Theoreticalframework} we carefully explain the theoretical framework. In Section~\ref{Results} we will then compare our NLO calculations with several contemporary proton PDFs against the ATLAS $13\,{\rm TeV}$ data \cite{measurementforcomparison} which comprises the most precise LHC measurement at the moment. In particular, we will point out the usefulness of the differential $W^+\overline{c}/W^-c$ ratios which can be measured very precisely and which prove to make a quantitative difference between the considered selection of PDFs. We also study the role of intrinsic charm quarks in the ATLAS measurement. Finally, we make some predictions for the processes of Eq.~\eqref{eq: p p > W D} in proton-lead ($p$Pb) collisions, discussing the feasibility of conducting an experiment to measure these observables.

\section{Theoretical framework}
\label{Theoreticalframework}

\begin{figure}[b!]
    \centering
    \begin{subfigure}{0.48\textwidth}
        \centering
        \includegraphics{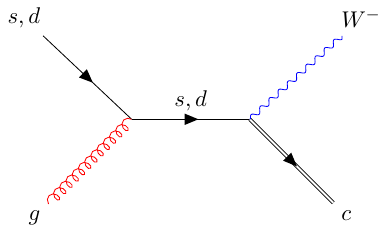}
        \caption{$s$ channel}
    \end{subfigure}
    \hspace*{0.2cm}
    \begin{subfigure}{0.45\textwidth}
        \centering
        \includegraphics{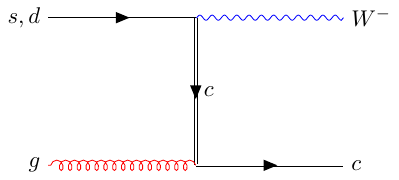}
        \caption{$t$ channel}
    \end{subfigure}
    \caption{Leading-order Feynman diagrams in the $p + p \rightarrow W^- + c + X$ process.}
    \label{fig: leading order W-c}
\end{figure}

\begin{figure}[t!]
    \centering
    \begin{subfigure}{0.45\textwidth}
        \centering
        \includegraphics{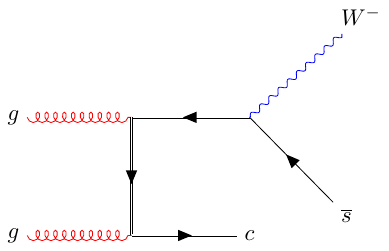}
        \caption{}
        \label{fig: gg channel}
    \end{subfigure}
    \begin{subfigure}{0.45\textwidth}
        \centering
        \includegraphics{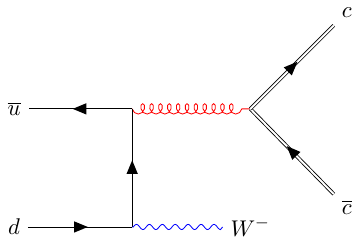}
        \caption{}
        \label{fig: OS-SS example}
    \end{subfigure}
    \caption{a) An example of a $gg$ channel subprocess introduced at NLO. b) An example of a diagram which cancels in the OS-SS subtraction.}
\end{figure}

Consider the following two processes:
\begin{align}
    \label{eq: reaction formula cW}
    p + p & \rightarrow W^- (\rightarrow l^- + \Bar{\nu}_l) + c + X \,, \\
    p + p & \rightarrow W^+ (\rightarrow l^+ + \nu_l) + \Bar{c} + X \,. \nonumber 
\end{align}
At leading order (LO) in perturbation theory, the partonic subprocess for the $W^-$ case can be represented by the Feynman diagrams shown in Figure \ref{fig: leading order W-c}. Of the down-type quarks in the initial state, the strange quark has the largest \textit{Cabibbo–Kobayashi–Maskawa} (CKM) matrix element at the $W$ boson vertex. This compensates for the fact that the $d/\Bar{d}$ PDFs tend to be larger than $s/\Bar{s}$. Due to the very small CKM matrix element the contributions from $b$-quark initial states are negligible. In fact, at LO the $sg$ initial state dominates the cross section. At NLO, new channels with $gg$ initial states appear: an example is shown in Figure \ref{fig: gg channel}. It turns out that these new channels have only a small contribution, leaving the $sg$ channel to dominate also at NLO.

Due to color confinement, the processes \eqref{eq: reaction formula cW} cannot be measured directly. The produced charm quark can be indirectly measured either through the jet of particles it produces or, more inclusively, by tagging a specific hadron from the jet. Jet measurements are widely used in PDF studies and the $W^\pm + {\rm charm \ jet}$ process is known up to NNLO in perturbative QCD \cite{Czakon:2020coa}. In this approach, however, modeling beyond collinear factorization is needed to account for the contributions of multi-parton interactions and broadening of parton-level jets. To suppress such non-perturbative effects, the transverse momentum of the jet has to be typically bounded rather strictly from below, decreasing the statistics of the process \cite{Dasgupta:2007wa}. 
To avoid this modeling and also consider a larger kinematic region, one can instead identify the quark by reconstructing a charmed hadron produced from it. For $c$ and $\Bar{c}$, the mesons $D^{(*)\pm}$ fit this role. The ATLAS experiment \cite{measurementforcomparison} tagged these mesons, i.e. measured the processes in Eq.~\eqref{eq: p p > W D}. In this article, we match our calculational setup to this experiment.

\begin{figure}[htb!]
    \centering
    \begin{subfigure}{0.45\textwidth}
        \centering
        \includegraphics{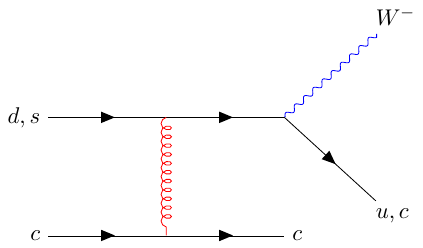}
        \caption{}
    \end{subfigure}
    \begin{subfigure}{0.45\textwidth}
        \centering
        \includegraphics{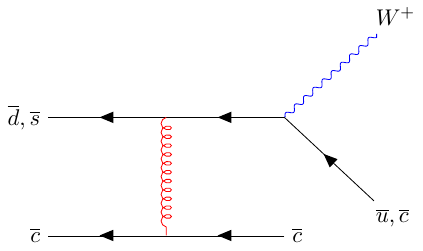}
        \caption{}
    \end{subfigure}
    \caption{Examples of diagrams with $c/\Bar{c}$ in the initial state contributing to the production of a) $W^- + c$ and b) $W^+ + \Bar{c}$ at NLO in perturbation theory. The diagrams arising from changing $c\leftrightarrow \Bar{c}$ in the incoming charm line cancel the ones shown here under the OS-SS subtraction if $c = \Bar{c}$ in PDFs.}
    \label{fig: c/cbar initial states}
\end{figure}

The ATLAS measurements correspond to the combinations of cross sections,
\begin{equation*}
    \sigma_\text{OS-SS} (W^\mp D^{(*)\pm}) = \sigma (W^\mp D^{(*)\pm}) - \sigma(W^\mp D^{(*)\mp}).
\end{equation*}
Here, the "same-sign" (SS) events are subtracted from the "opposite-sign" (OS) events, in what is coined the "OS-SS" subtraction. The reasoning behind such a subtraction is that it removes a lot of background processes and partonic channels with no strange quark in the initial state, thus increasing the sensitivity of the observables to the $s$ and $\Bar{s}$ PDFs. An example of a diagram that is canceled in the subtraction is shown in Figure \ref{fig: OS-SS example}. If $c = \Bar{c}$ is assumed, the class of canceled diagrams also includes those with $c$ or $\Bar{c}$ in the initial state: examples of these diagrams are shown in Figure \ref{fig: c/cbar initial states}. Thus, the $c/\Bar{c}$ PDFs can only contribute to the measured cross sections through a possible asymmetry $c - \Bar{c} \neq 0$. Also the contributions of double-parton scattering, in which the D-meson and the $W$ are created in independent partonic collisions, will cancel in the OS-SS subtraction.

\subsection{Cross sections in GM-VFNS}
\label{sec: frag}

Taking the case of $W^-D^+$ production as an example, the cross sections differential in the tranverse momentum of the meson, $p_{T, D}$, the
pseudorapidity of the meson, $\eta_D$, and the pseudorapidity of the charged lepton, $\eta_l$  
can be calculated in the fixed-flavour-number scheme by,
\begin{align}
    \label{eq: fragmentation main equation three variables}
    \frac{d\sigma(W^-D^+)}{dp_{T, D} d\eta_D d\eta_l} & = \int_{z_\text{min}}^1 dz \int dp_{T, c} \int d\eta_c \, D_{c\rightarrow D^+} (z) \frac{d\sigma(W^-c,\mu_\text{ren}, \mu_\text{fact})}{dp_{T, c} d\eta_c d\eta_l}  \notag \\[10pt]
    & \times \delta \Big[p_{T, D} - p_{T, D} \Big(p_{T, c}, \eta_c, z\Big)\Big] \delta \Big[\eta_D - \eta_D \Big(p_{T, c}, \eta_c, z\Big)\Big].
\end{align}
The lower bound of the $z$ integral, $z_{\rm min}$, depends on both $p_{T, D}$ and $\eta_D$, as will be discussed below. The differential cross section on the right-hand side of the equation is for the process $p + p \rightarrow W^-(\rightarrow l^- + \Bar{\nu}_l) + c + X$ which we in this work evaluate by the publicly available MCFM code \cite{MCFM_1, MCFM_2, MCFM_3}.
This partonic cross section is convoluted with the scale-independent fragmentation function $D_{c\rightarrow D^+}(z)$ to yield the cross section where the final state quark is replaced by the meson. The kinematics of the quark and the meson are related through the assumption of \textit{collinear fragmentation}, which means that the quark and the meson are collinear, and by defining a \textit{fragmentation variable} $z$.
Denoting the 4-momenta of the quark, the meson and the incoming protons as $p_c$, $p_D$, $p_1$ and $p_2$ respectively, a possible definition of $z$ reads
\begin{equation}
    \label{eq: z definition KKKS08}
    z_+ \equiv \frac{p_D \cdot (p_1 + p_2)}{p_c \cdot (p_1 + p_2)} \xrightarrow{\text{CM frame}} \frac{E_D}{E_c}.
\end{equation}
As indicated, in the center-of-mass (CM) frame of the colliding hadrons the fragmentation variable is given by the ratio between the energies of the meson and the quark. We choose to use a slightly different definition employed e.g. in Ref.~\cite{Helenius:2023wkn}, which reads
\begin{equation}
    \label{eq: z definition}
    z_- \equiv \frac{p_D \cdot (p_1 - p_2)}{p_c \cdot (p_1 - p_2)} \xrightarrow{\text{CM frame}} \frac{p_{T, D}}{p_{T, c}}.
\end{equation}
With this alternative definition, in the CM frame of the colliding hadrons $z$ is given by the ratio between the transverse momenta of the meson and the quark, $p_{T, D}$ and $p_{T, c}$. This definition of $z$ avoids pathological behavior arising at low transverse momenta when using the other definition of $z=z_-$ 
\cite{Helenius:2023wkn}.
However, in the analysis done in this article the transverse momentum of the meson is large enough that this problem does not become relevant. Indeed, in Section \ref{sec: pp results} we show that the numerical differences between these two definitions of $z$ in the physical observables is negligible for $p_{T, D} \gtrsim 8 \,{\rm GeV}$ which is the lower bound considered in the present work. The benefit of $z_+$ is that it makes the expressions that follow slightly simpler. Note that the two definitions are equal in the massless limit.

Having decided on the definition of $z$, we can discuss its integration limits in equation \eqref{eq: fragmentation main equation three variables}. The upper boundary of $1$ corresponds to the meson carrying $100\%$ of the quark's transverse momentum. The lower boundary corresponds to the case where the transverse momentum of the quark is at its maximum.
The assumption of collinear fragmentation allows us to write $z$ in terms of the ratio between the 3-momenta of the meson and the quark, and we get
\begin{equation}
    \label{eq: zmin}
    z_\text{min} = \frac{|\overline{p}_D|}{|\overline{p}_c|_\text{max}} = \frac{2\sqrt{s}}{s - m_c^2} \frac{p_{T, D}}{\sin[2\arctan(e^{-\eta_D})]},
    \end{equation}
where $\sqrt{s}$ is the CM energy of the collision. 

Before implementing equation \eqref{eq: fragmentation main equation three variables} numerically, we want deal with the Dirac delta functions. Due to collinear fragmentation $\eta_D (p_{T, c}, \eta_c, z) = \eta_c$ and thus
\begin{equation*}
    \int d\eta_c \, \delta \Big[\eta_D - \eta_D \Big(p_{T, c}, \eta_c, z\Big)\Big] = 1,
\end{equation*}
and from the definition of the fragmentation variable
\begin{equation*}
    \int dp_{T, c} \, \delta \Big[p_{T, D} - p_{T, D} \Big(p_{T, c}, \eta_c, z\Big)\Big] = \frac{1}{z}.
\end{equation*}
Therefore equation \eqref{eq: fragmentation main equation three variables} simplifies to
\begin{align}
    \label{eq: differential cross section only z integral}
    \frac{d\sigma(W^-D^+)}{dp_{T, D} d\eta_D d\eta_l} =& \int_{z_\text{min}}^1 \frac{dz}{z} D_{c\rightarrow D^+} (z) \frac{d\sigma(W^-c,\mu_\text{ren}, \mu_\text{fact})}{dp_{T, c} d\eta_c d\eta_l},
\end{align}
where
$p_{T, c} = p_{T, D}/z$ and $\eta_c = \eta_D$. In practice, we restrict the lower bound of the $z$ integral given in equation \eqref{eq: zmin} by not letting it go lower than $z = 0.05$, which is the lower boundary of the applied KKKS08 fit of FFs \cite{KKKS08}. However, in Section \ref{sec: pp results} we show that this has a vanishing effect on the cross sections.

\begin{figure}[b!]
    \centering
    \includegraphics{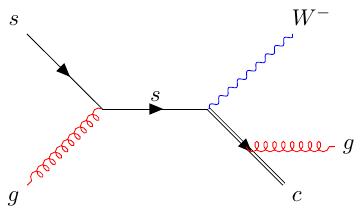}
    \caption{An example of a diagram included in the NLO calculation of the process $p + p \rightarrow W^- + c + X$, where the radiated gluon can be emitted collinearly with the quark, producing a large logarithm.}
    \label{fig: gluon radiation}
\end{figure}

The expression above now corresponds to a fixed-order prediction in which the NLO partonic cross section given by MCFM effectively involves logarithmic terms of the form $\log (p_{T, c} / m_c)$ \cite{Giele:1995kr}. These logarithms originate from collinear radiation of gluons from final-state charm quarks, an example of which is shown in Figure \ref{fig: gluon radiation}. In GM-VFNS, these logarithms are summed to the scale dependent FFs $D_{i\to h}(z, \mu_{\rm frag}^2)$ which follow the usual time-like $\overline{\rm MS}$ DGLAP equations. To first order in $\alpha_s$,
 \cite{dQQ,Kretzer:1998nt}
\begin{equation}
\label{eq:scale_dep_ff}
\begin{aligned}
    D_{c\rightarrow D^+}(z, \mu_{\rm frag}^2)&=D_{c\to h}(z)+\frac{\alpha_s(\mu_{\rm ren}^2)}{2\pi}C_F\int_z^1 \frac{dl}{l} D_{c\rightarrow D^+}(\frac{z}{l}) \\
    &\phantom{=} \ \times\left[\frac{1+l^2}{1-l}\left(\log\frac{\mu_{\rm frag}^2}{m_c^2}-1-2\log(1-l)\right)\right]_++\mathcal{O}(\alpha_s^2),
\end{aligned}
\end{equation}
where $C_f = 4/3$ and the plus distribution is to be understood as
\begin{equation*}
    \int_0^1 dx [g(x)]_+ f(x) = \int_0^1 dx g(x) \left[f(x) - f(1)\right] \,.
\end{equation*}
Inverting this relation gives,
\begin{equation}
\begin{aligned}
    D_{c\rightarrow D^+}(z)&=D_{c\rightarrow D^+}(z, \mu_{\rm frag}^2)-\frac{\alpha_s(\mu_{\rm ren}^2)}{2\pi}C_F\int_z^1 \frac{dl}{l} D_{c\rightarrow D^+}(\frac{z}{l}, \mu_{\rm frag}^2) \\
    & \times\left[\frac{1+l^2}{1-l}\left(\log\frac{\mu_{\rm frag}^2}{m_c^2}-1-2\log(1-l)\right)\right]_++\mathcal{O}(\alpha_s^2).
\end{aligned}
\end{equation}
Using this in eq.~\eqref{eq: differential cross section only z integral} replaces the scale-independent FF with the scale-dependent one and gives an additional subtraction term
\begin{align}
    \label{eq:lopullinen}
    \frac{d\sigma(W^-D^+)}{dp_{T, D} d\eta_D d\eta_l} = & \int_{z_\text{min}}^1 \frac{dz}{z} D_{c\rightarrow D^+} (z, \mu_\text{frag}) \frac{d\sigma(W^-c,\mu_\text{ren}, \mu_\text{fact})}{dp_{T, c} d\eta_c d\eta_l} \\
     - & \int_{z_\text{min}}^1 \frac{dz}{z} L_{c\rightarrow D^+}(z, \mu_\text{frag}) \frac{d\sigma_{\rm LO}(W^-c,\mu_\text{ren}, \mu_\text{fact})}{dp_{T, c} d\eta_c d\eta_l} \nonumber
\end{align}
where $d\sigma_{\rm LO}$ refers to the leading-order partonic cross section and
\begin{align}
    \label{eq: subtraction FF}
    L_{c\rightarrow D^+}(z, \mu_\text{frag}) & = \frac{\alpha_s(\mu_{\rm ren}^2)}{2\pi}C_F\int_z^1 \frac{dl}{l} D_{c\rightarrow D^+}(\frac{z}{l}, \mu_{\rm frag}^2) \\
    & \times\left[\frac{1+l^2}{1-l}\left(\log\frac{\mu_{\rm frag}^2}{m_c^2}-1-2\log(1-l)\right)\right]_++\mathcal{O}(\alpha_s^2) \nonumber \,.
\end{align}
Following the logic of Ref.~\cite{Guzzi:2024can}, the function $L_{c\rightarrow D^+}$ could be called a ``subtraction FF''. The role of the subtraction term is to remove the logarithmic term which is already taken into account by the scale-dependent fragmentation function. The non-logarithmic terms are necessary for the calculation to match the zero-mass $\overline{\rm MS}$ scheme in the $m_c \rightarrow 0$ limit. In our calculation our default is to keep all the non-physical scales equal,
\begin{equation*}
    \mu_\text{fact} = \mu_\text{ren} = \mu_\text{frag} = M_W.
\end{equation*}
As a result, we will not need to supply an additional logarithmic terms $\propto \alpha_s^2 \log(\mu_\text{ren}/\mu_\text{fact})$ to account for the difference between the decoupling and $\overline{\rm MS}$ renormalization schemes \cite{Bevilacqua:2021ovq}. However, we will still consider variations around the default scale by varying all three scales uniformly up and down by a factor of two. We also study an alternative choice $\mu_\text{frag} = p_{T, D}$ for the fragmentation scale inspired by the logarithmic dependence of the partonic cross sections on the charm transverse momentum.

\subsection{Computational setup}
\label{sec: specifics of the calculation}
The calculation of $W+c$ cross sections is done mainly with MCFM 10.3 \cite{MCFM_1, MCFM_2, MCFM_3}. The only exception comes when considering charm-quark initial states, where MadGraph5\_aMC@NLO \cite{MadGraph} is used. The two programs were cross checked against each other at LO. For proton-lead collisions, the MCFM source code had to be modified to allow different PDFs for the two colliding objects. This modification was tested at LO against an independent implementation \cite{Alanko2025}. We evalute these parton-level cross sections in bins of $p_{T, c}$, $\eta_c$ and $\eta_l$, which are then convoluted with the scale-dependent FFs as indicated in Eq.~(\ref{eq:lopullinen}).

The PDFs used in this article are available through the LHAPDF library \cite{LHAPDF}. The PDFs considered are CT18ANLO \cite{Hou:2019efy}, MSHT20nlo\_as118 \cite{MSHT20PDF} and NNPDF40\_nlo\_as\_01180 \cite{NNPDF40PDF} with and without perturbative charm (pch) for $pp$ collisions, and nNNPDF30\_nlo\_as\_\\01180 \cite{AbdulKhalek:2022fyi} and EPPS21nlo \cite{EPPS21} for $p$Pb collisions. All PDF errors shown in this article are $68\%$ confidence level (C.L.). 
For FFs our default choice is KKKS08 \cite{KKKS08} fitted to OPAL data \cite{OPAL:1996ikk,OPAL:1997edj} in $e^+e^-$ collisions, but also the variations induced by using the global data set, as well as by using a different fit to the OPAL data, SMSKA19 \cite{SMSKA19}, are examined. The OPAL data correspond to CM energy $\sqrt{s} = M_Z$ which is way larger than the charm-quark mass and e.g. the scheme ambiguities in treating the charm-quark mass effects should be rather irrelevant. In this sense, we consider the FF extraction based on OPAL data more robust than the global one which also includes low-energy BELLE \cite{Belle:2005mtx} and CLEO \cite{CLEO:2004enr} data from $e^+e^-$ collisions where the mass effects are presumably much more relevant. Indeed, there is a known mismatch between measured energy dependence and the one predicted by the NLO DGLAP evolution \cite{Cacciari:2005uk,Bonino:2023icn}. It was recently demonstrated \cite{Cacciari:2024kaa} that the discrepancy can be alleviated by an appropriate treatment of heavy-quark threshold effects in resummation of mass and soft logarithms. This mostly affects the low-energy cross sections which consolidates the use of OPAL-data fitted FFs. The SMSKA19 fit is only provided at the initial scale of $\mu_\text{frag} = \sqrt{18.5} \, \text{GeV}$, and we use the EKO \cite{eko} program to perform the evolution to $\mu_\text{frag} = M_W$.

Our calculations use the following parameter values. The Fermi coupling is taken as
\begin{equation*}
    G_\mu = 1.16639 \times 10^{-5} \,\text{GeV}^{-2},
\end{equation*}
while the strong coupling constant $\alpha_s$ is given by LHAPDF to match the value in each PDF set. The $W$ boson mass is set to
\begin{equation*}
    \quad M_W = 80.385 \, \text{GeV} \,,
\end{equation*}
and for the charm quark mass we use the values used in the PDF sets, which are
\begin{equation*}
m_c = 
\begin{cases}
        1.3 \, \text{GeV} \ \ \, (\text{CT18A/EPPS21})\\
        1.4 \, \text{GeV} \ \ \, (\text{MSHT20}) \\
        1.51 \,\text{GeV} \ (\text{NNPDF4.0/NNPDF3.0})
\end{cases}        
    .
\end{equation*}
The decay width of the $W$ boson is
\begin{equation*}
    \Gamma_W = 2.093 \, \text{GeV},
\end{equation*}
and the relevant CKM matrix elements are
\begin{equation*}
    V_{cd} = 0.222 \quad \text{and} \quad V_{cs} = 0.975.
\end{equation*}
Our analysis for both $pp$ and $p$Pb collisions applies the kinematic cuts of the ATLAS experiment \cite{measurementforcomparison}, outlined in Table \ref{tab: cuts}. The collision energies are $\sqrt{s} = 13$ TeV in $pp$ collisions and $\sqrt{s} = 8.5$ TeV in $p$Pb collisions.
\begin{table}[t!]
\def\arraystretch{1.8}
\centering
    \caption{Kinematic cuts used in the ATLAS analysis \cite{measurementforcomparison}.}
    \begin{tabular}{ c | c }
        \hline
        Quantity & Cut\\
        \hline
        Charged lepton $p_T$ & $> 30$ GeV\\
        \hline
        Charged lepton  $\lvert \eta \rvert$  & $< 2.5$\\
        \hline
        $D^{(*)\pm}$ $p_T$ & $> 8$ GeV\\
        \hline
        $D^{(*)\pm}$ $\lvert \eta \rvert$ & $< 2.2$\\
        \hline
    \end{tabular}
    \label{tab: cuts}
\end{table}

\section{Results}
\label{Results}

In the following, we will go over numerical results calculated within the 
framework explained in Section \ref{Theoreticalframework}. 
In the case of $pp$ collisions we compare our predictions with the experimental data from ATLAS \cite{measurementforcomparison}, and then use the PDF reweighting techniques to study the implications of the ATLAS data on PDFs. 
We will also discuss the stability of the results to various choices made in the calculation, as well as the effects of intrinsic charm and the numerical relevance of the subtraction term in equation \eqref{eq:lopullinen}. For the $p$Pb collisions we provide theoretical predictions and estimate the feasibility of measuring $W^\mp D^{(*)\pm}$ cross sections.  

\subsection{Proton-proton collisions}
\label{sec: pp results}

\begin{figure}[b!]
    \centering
    \includegraphics[width=0.48\linewidth]{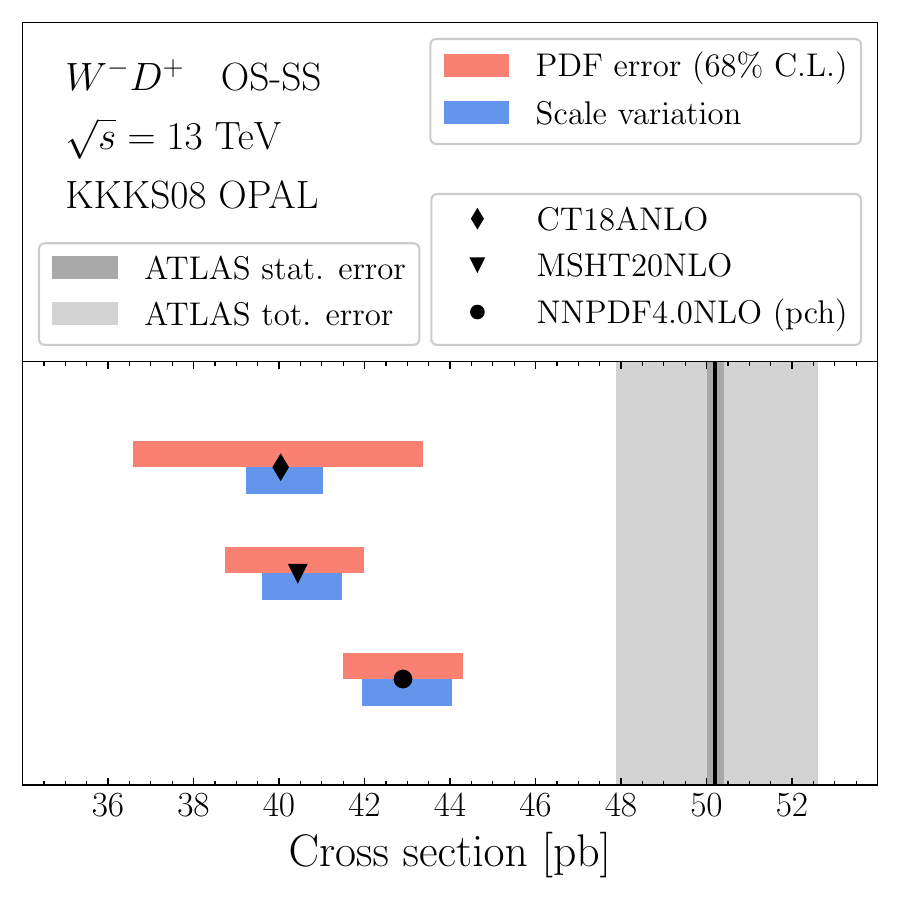}
    \includegraphics[width=0.48\linewidth]{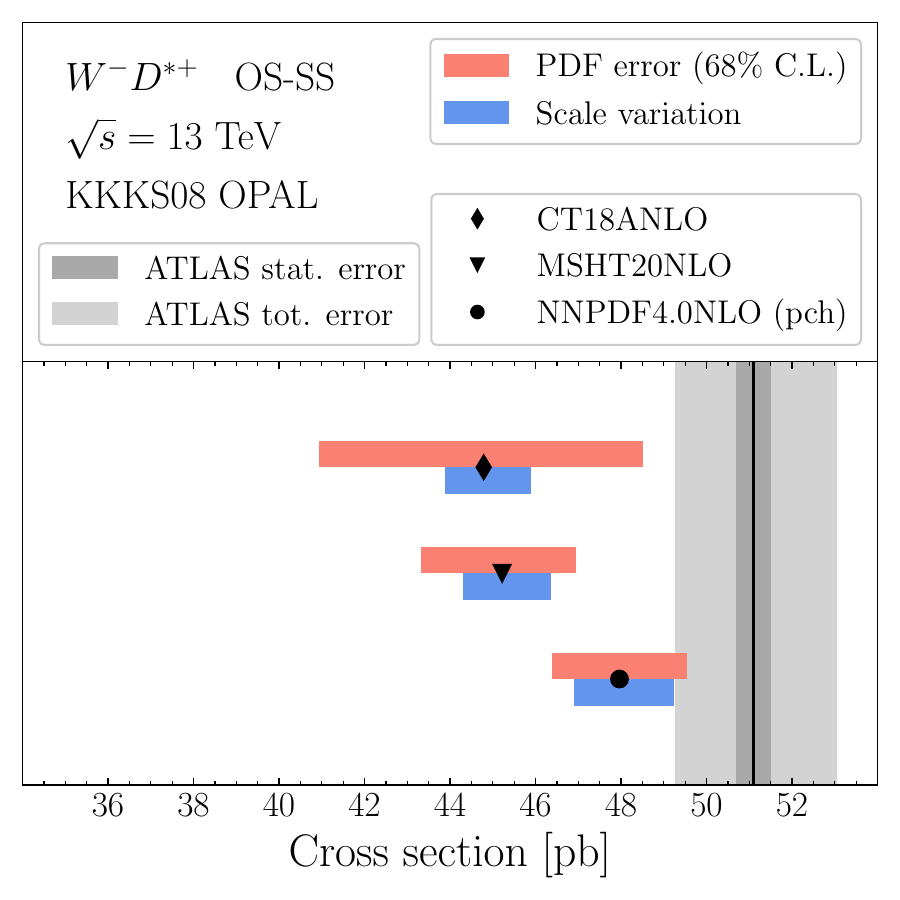}
    \includegraphics[width=0.48\linewidth]{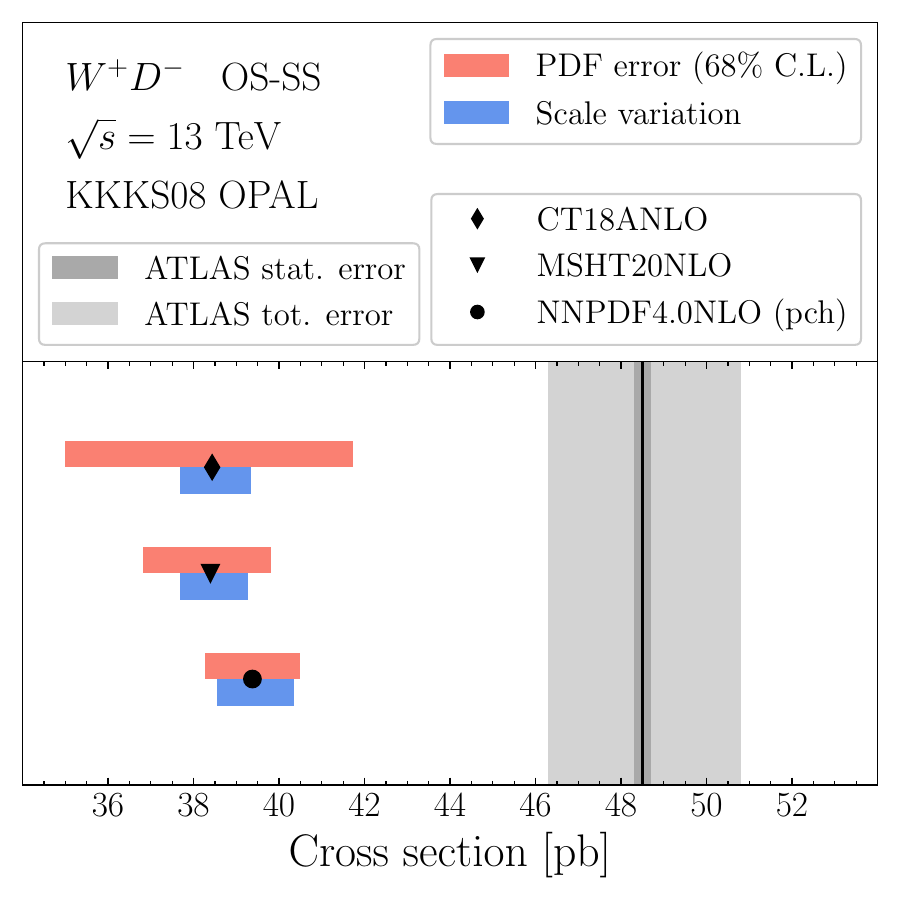}
    \includegraphics[width=0.48\linewidth]{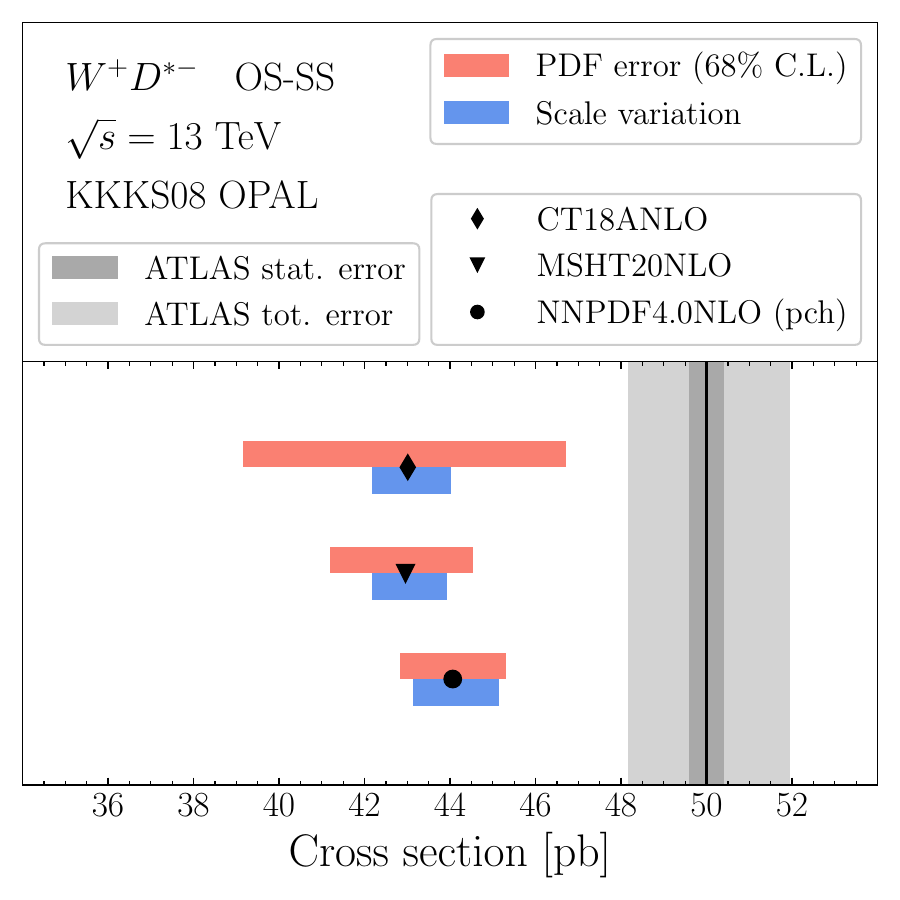}
    \caption{Integrated NLO cross sections for the processes $pp\rightarrow W^\mp D^{(*)\pm} + X$ compared with the ATLAS measurements \cite{measurementforcomparison}.}
    \label{fig: pp integrated}
\end{figure}

Figure~\ref{fig: pp integrated} presents cross sections for the four processes $p + p\rightarrow W^\mp + D^{(*)\pm}+ X$ integrated over the kinematic cuts of Table~\ref{tab: cuts}.
The three different PDF sets yield predictions which agree for the most part, the biggest difference being between NNPDF and the two others in $W^-D^{(*)+}$ production. The predictions are, however, quite significantly below the ATLAS data even when accounting for the scale variations and PDF uncertainty, especially in the cases involving $D^{\pm}$. This tension is known to reduce greatly with the inclusion of NNLO corrections: zero-mass  calculation shows a correction of $\sim 15\%$ upwards for all processes which is nearly constant in the kinematic region considered \cite{NNLOcalculation}. The FF uncertainties are not shown here -- the FF dependence will be discussed below in more detail. The calculated production ratio between the two meson types $[\sigma(W^\mp D^\pm) / \sigma(W^\mp D^{*\pm})]_{\rm NLO} \sim 0.9$ is smaller than what the data indicate,  $[\sigma(W^\mp D^\pm) / \sigma(W^\mp D^{*\pm})]_{\rm exp} \sim 0.97 \ldots 0.98$ with the central values of the data. At the same time, this experimental value is roughly consistent with the relative charm-hadron production fractions measured at LEP, see e.g.  \cite{dEnterria:2026tuz} for a concise review. This implies that some modifications to KKKS08 charm-hadron FFs would be needed to better reproduce the data systematics. We also note that the measured relative abundances of $D^\pm$ and $D^{*\pm}$ here do not seem to follow the trend seen in the inclusive charm production in $pp$ collisions where it has been observed that the production of $D^\pm$ becomes actually more likely than that of $D^{*\pm}$ along with an increased rate of baryon production \cite{LHCb:2015swx,ALICE:2021dhb,ALICE:2023sgl}, although within the systematic uncertainties shown in Figure~\ref{fig: pp integrated} the evidence is not conclusive. These effects have been noticed to originate mostly from low values of $p_{T, D}$ (a few ${\rm GeV}$s) \cite{ALICE:2023sgl} and the cut $p_{T, D} > 8 \, {\rm GeV}$ imposed here may partly explain why the modified production fractions are not visible here. The inclusive production is also overwhelmed by the heavy-quark pair production i.e. the partonic processes there are very different than the ones relevant here. A common explanation attributes the different production fractions in $pp$ vs. $e^+e^-$ to a denser surrounding partonic environment which affects the fragmentation process. This explanation would thus seem to predict the low-$p_{T, D}$ production fractions to be different in $W^\pm + {\rm charm}$ processes as well. The $W^\pm + {\rm charm}$ processes in $pp$ should thus be relevant in understanding the modifications in production fractions of charmed hadrons and would motivate a measurement with a lower $p_{T, D}$ cut than the one used in the ATLAS measurement here. 

 \begin{figure}[t!]
    \centering
    \includegraphics[width=0.49\linewidth]{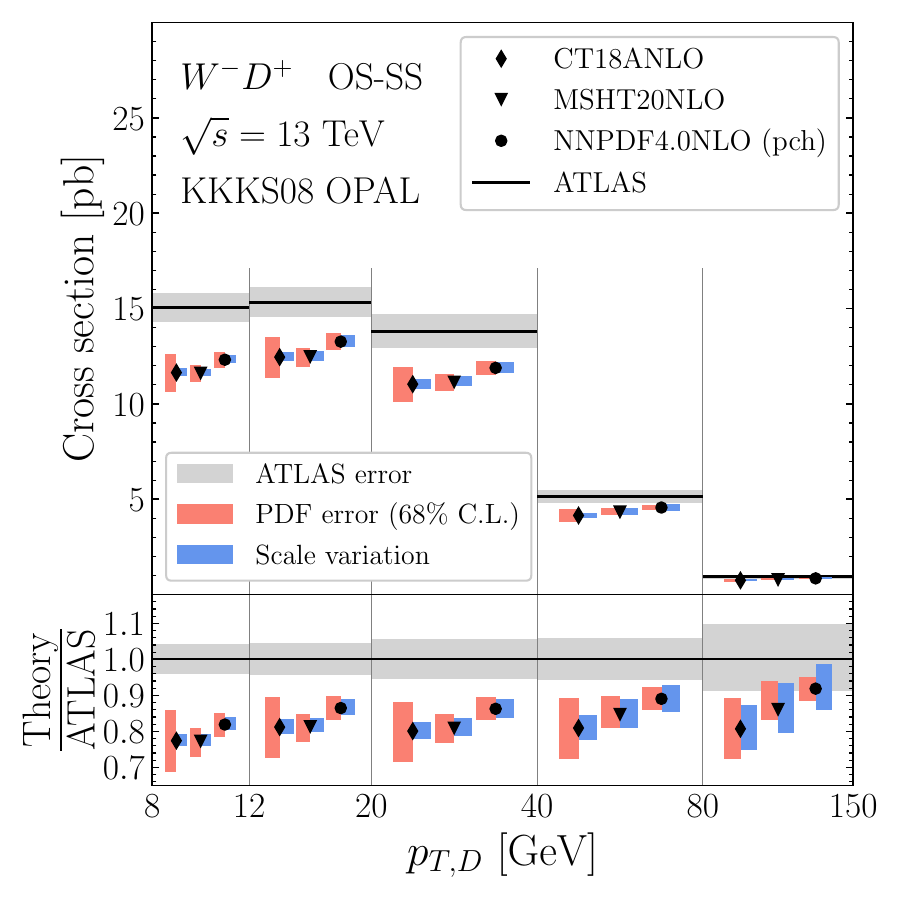}
    \includegraphics[width=0.49\linewidth]{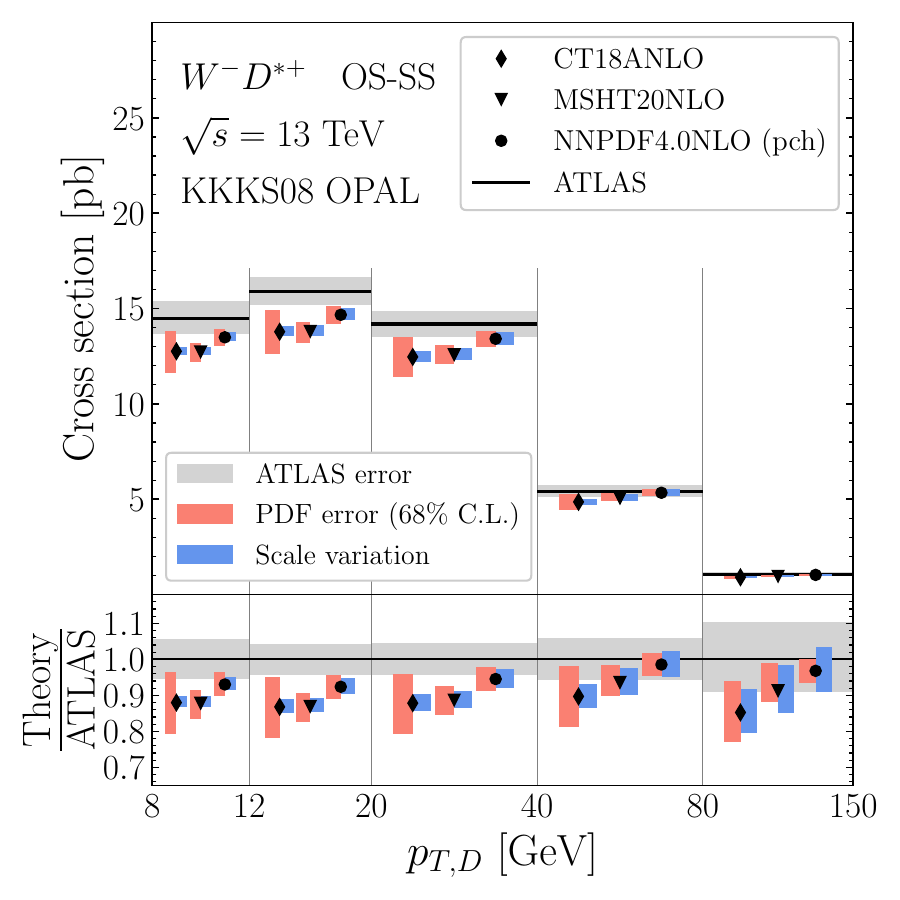}
    \includegraphics[width=0.49\linewidth]{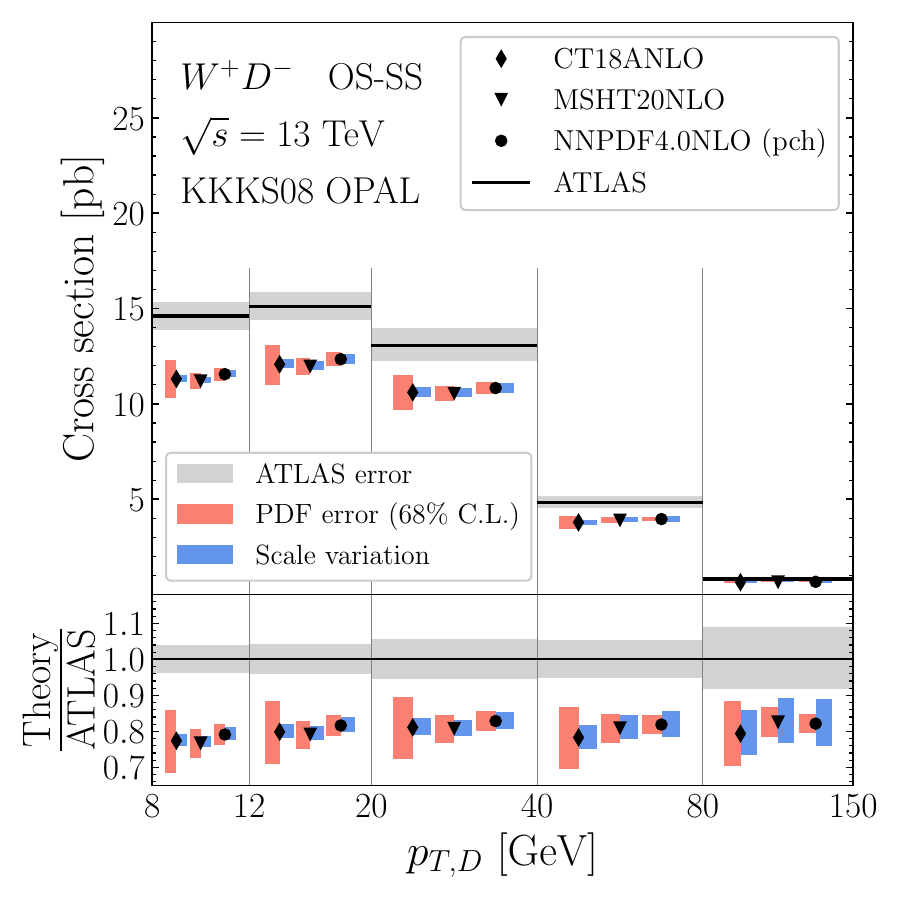}
    \includegraphics[width=0.49\linewidth]{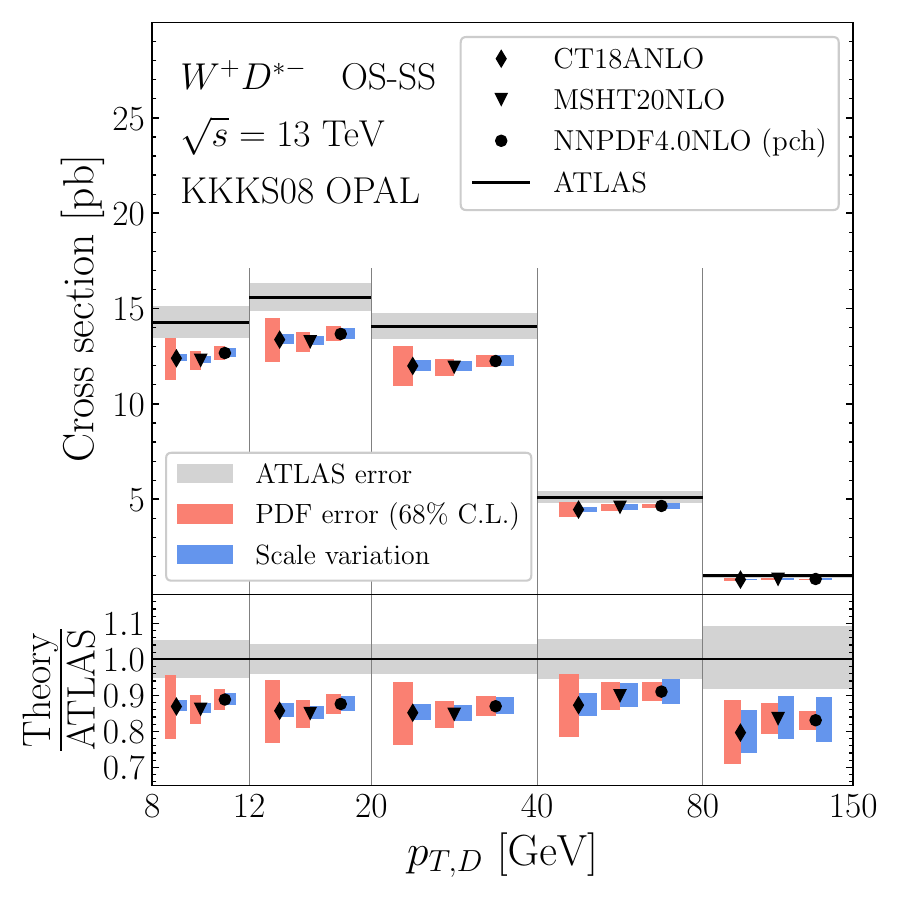}
    \caption{Bin-integrated cross sections for the processes $pp\rightarrow W^\mp D^{(*)\pm} + X$ as a function of transverse momentum of the meson compared with the ATLAS measurements \cite{measurementforcomparison}.}
    \label{fig: differential pTD}
\end{figure}

\begin{figure}[t!]
    \centering
    \includegraphics[width=0.49\linewidth]{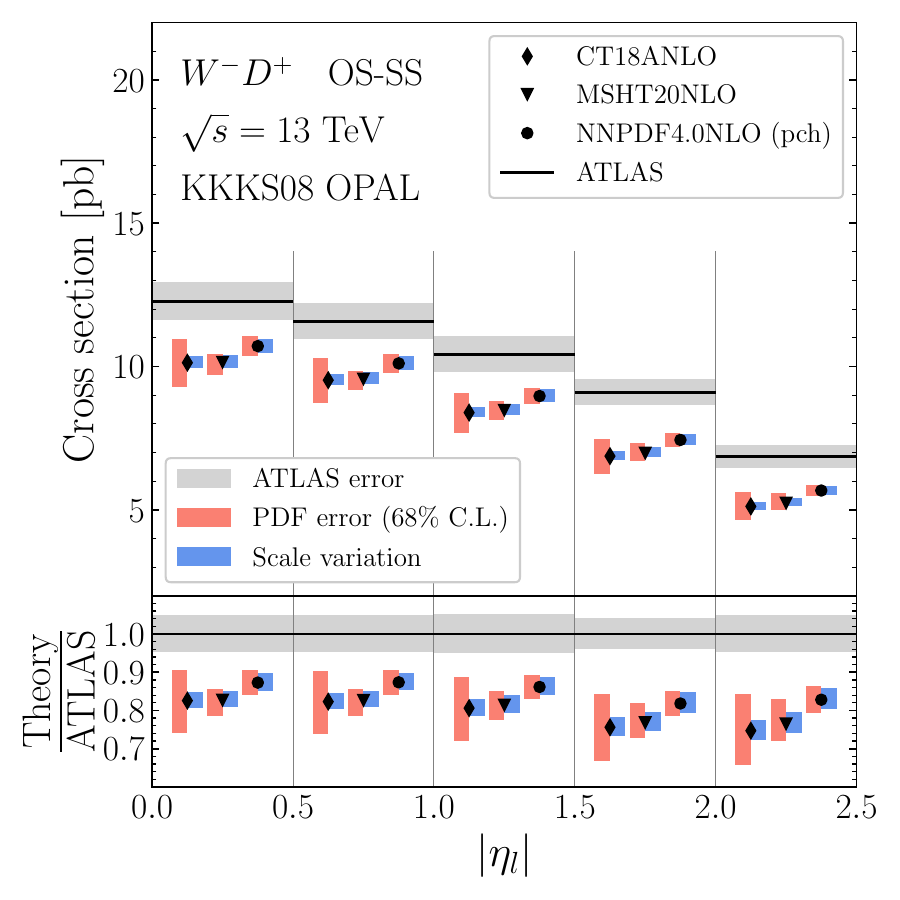}
    \includegraphics[width=0.49\linewidth]{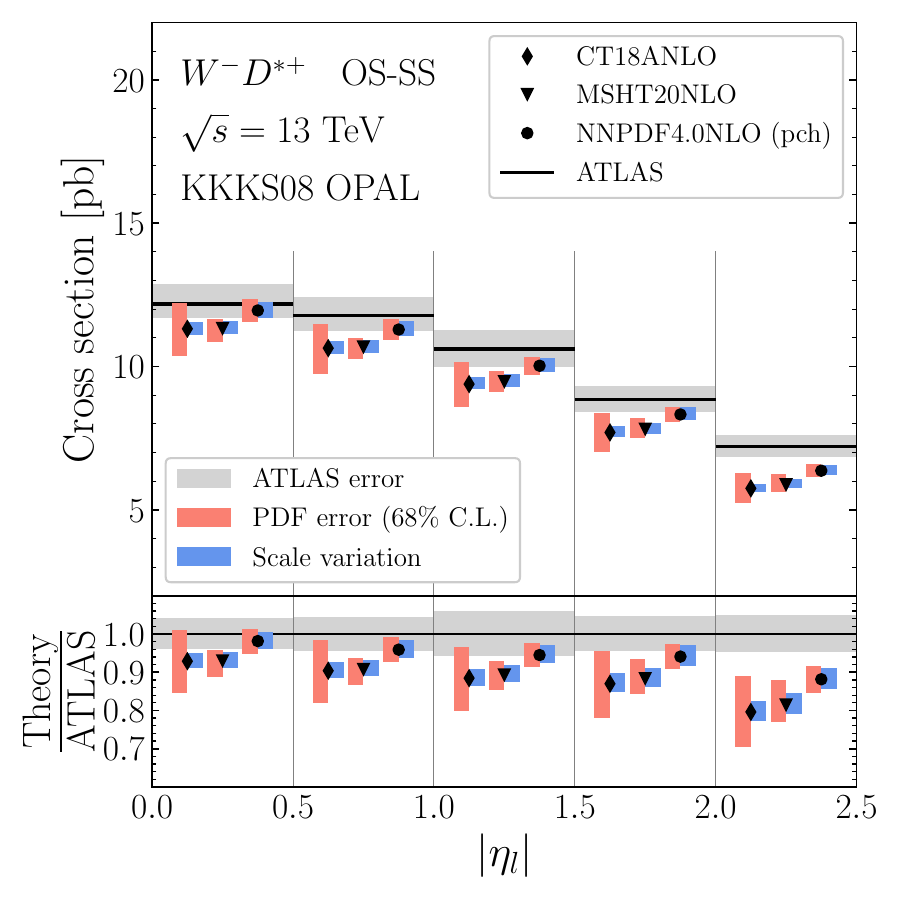}
    \includegraphics[width=0.49\linewidth]{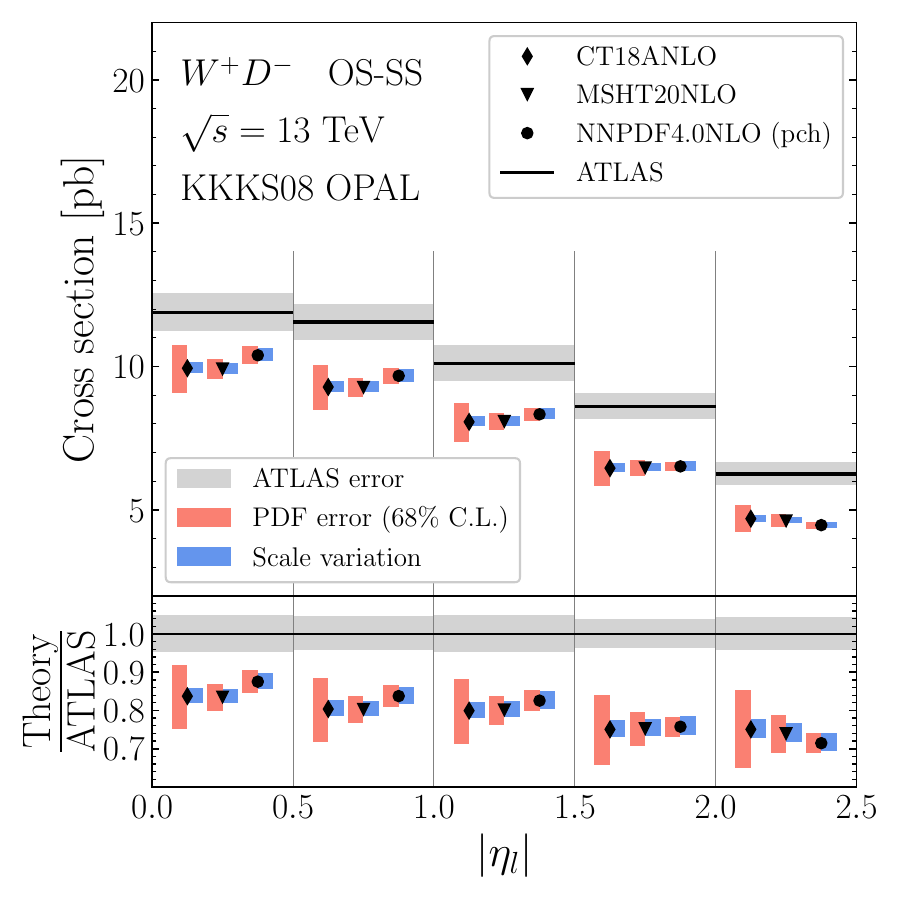}
    \includegraphics[width=0.49\linewidth]{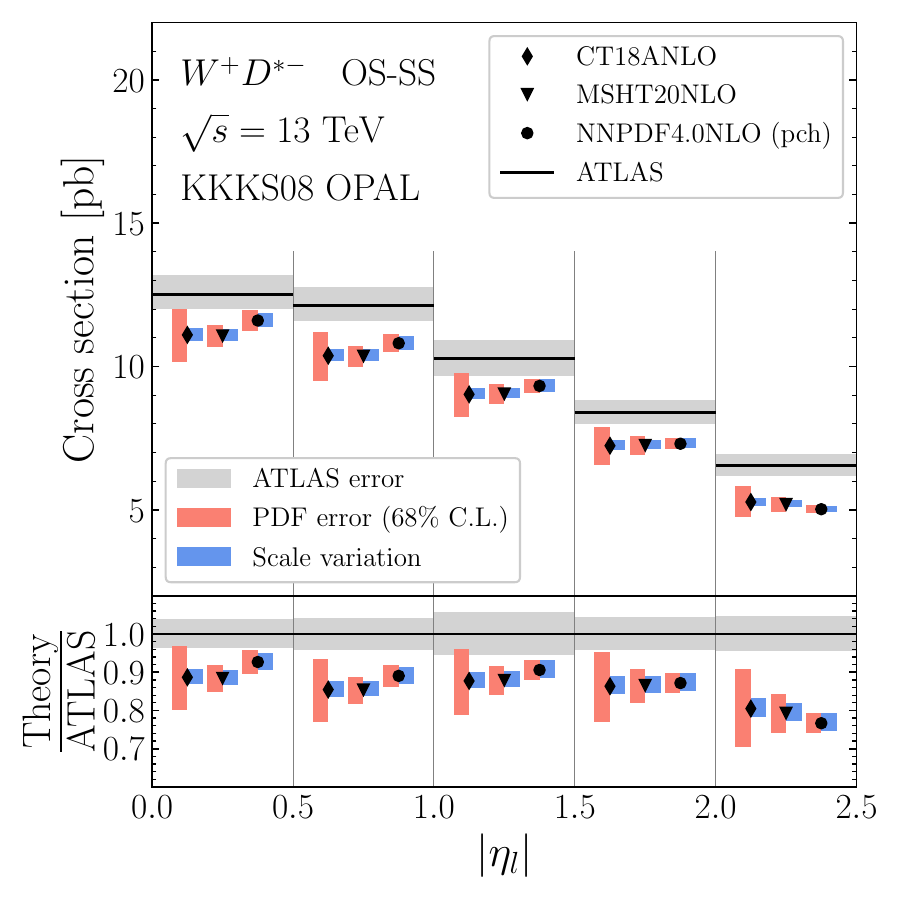}
    \caption{Bin-integrated cross sections for the processes $pp\rightarrow W^\mp D^{(*)\pm} + X$ as a function of psudorapidity of the meson compared with the ATLAS measurements \cite{measurementforcomparison}.}
    \label{fig: differential eta_lept}
\end{figure}

Figures \ref{fig: differential pTD} and \ref{fig: differential eta_lept} present the $p_{T, D}$- and $\eta_l$-differential cross sections, respectively, integrated over the bin widths. 
The differences between the data and the NLO predictions follow closely the pattern seen in Figure~\ref{fig: pp integrated} with some smallish 
kinematic dependence. These results are roughly compatible with the zero-mass results of Ref.~\cite{NNLOcalculation} which indicates that within the considered kinematic regime the residual charm-mass effects are rather weak. Interestingly, we note that the parton-level NLO calculations matched to a parton shower, presented in the ATLAS publication \cite{measurementforcomparison}, are in a better agreement with these differential data than the FF-based calculation presented here. As the perturbative accuracy is the same in both calculations, we attribute the differences to the way the hadronization is handled in these two approaches. However, the NNLO corrections are still known \cite{NNLOcalculation} to increase the cross sections by some 15\% which would then have a tendency to increase the parton-shower-matched cross sections above the measured values.

\begin{figure}[t!]
    \centering
    \includegraphics[width=0.49\linewidth]{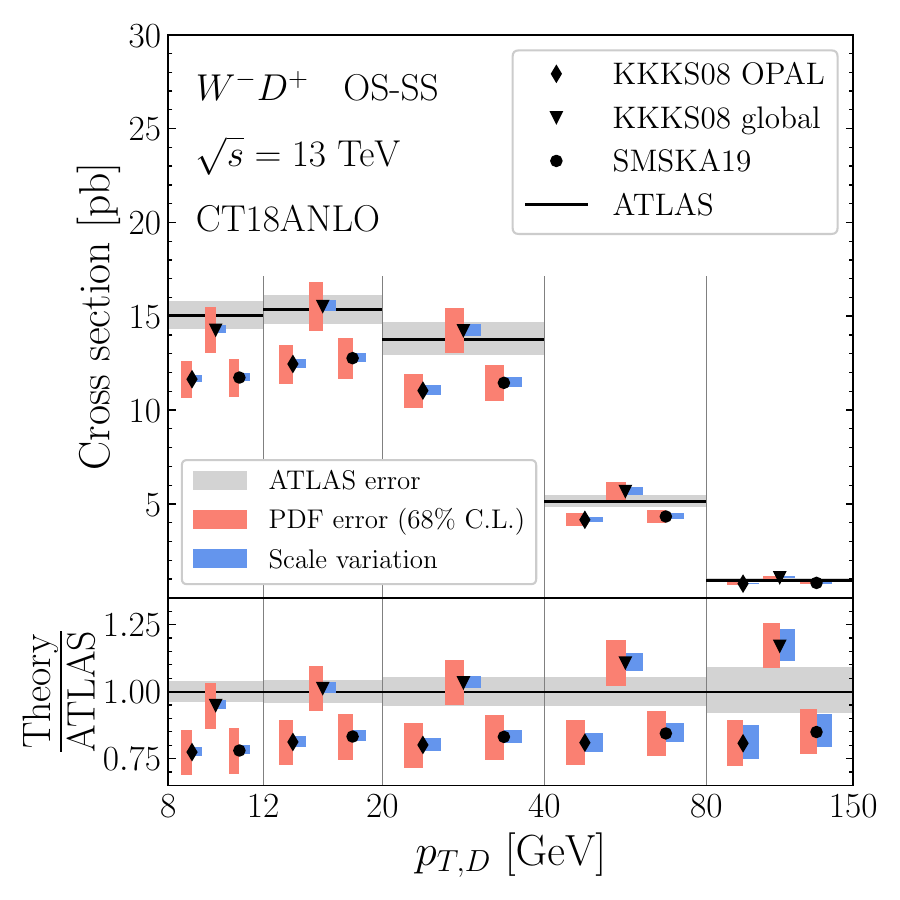}
    \includegraphics[width=0.49\linewidth]{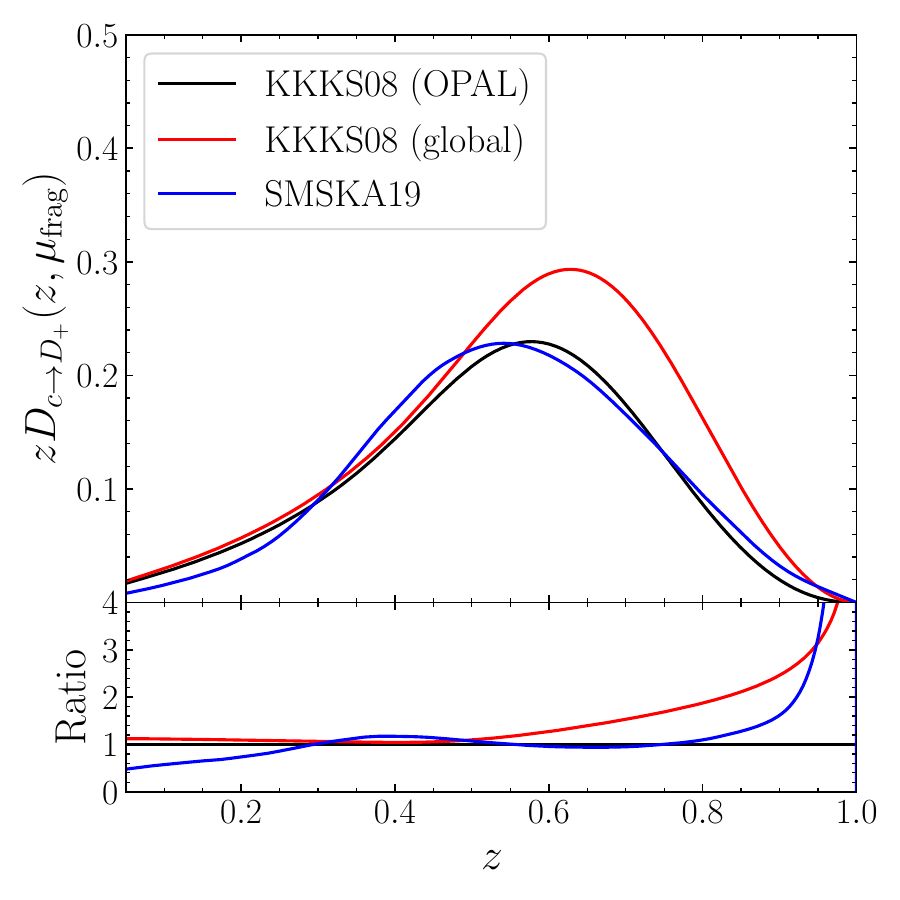}
    \caption{The dependence of the $p_{T, D}$ bin-integrated $W^-D^+$ cross section on the choice of FFs compared with the ATLAS measurements \cite{measurementforcomparison} (left) and the $z$ dependence of the corresponding FFs at $\mu_{\rm frag} = M_W$ (right). 
    }
    \label{fig: sigma pTD opal vs global}
\end{figure}

The sensitivity of our NLO predictions to the FFs is illustrated in the left-hand panel of Figure~\ref{fig: sigma pTD opal vs global} showing the $p_{T, D}$ dependent cross section for $D^+$ for three different sets of FFs: KKKS08 OPAL, KKKS08 global, and SMSKA19. While KKKS08 OPAL and SMSKA19 yield very similar results with theory-to-data ratio being nearly constant as a function of $p_{T, D}$, KKKS08 global yields clearly larger cross sections with a visible $p_{T, D}$ dependence in the theory-to-data ratio. This ordering can be understood from the right-hand panel of Figure~\ref{fig: sigma pTD opal vs global} which compares these three FFs at $\mu_{\rm frag} = M_W$. While KKKS08 OPAL and SMSKA19 are mutually more or less consistent between $0.3 < z < 0.8$, KKKS08 global deviates from these two at $z > 0.5$ leading to distinct behaviour that can be seen in the left-hand panel. The sensitivity of the $p_{T, D}$-differential cross sections to the $z$ dependence of the FFs can be explained by the behaviour of the partonic $p_{T, c}$ spectrum which decreases steeper as $p_{T, c}$ increases. This tends to suppress the contributions from $z < 1$ increasingly strongly as $p_{T, D}$, which sets the lower limit for the contributing $p_{T, c}$, increases. As a result, the large $p_{T, D}$ end is more sensitive to FFs at larger $z$. This also indicates that the studied $p_{T, D}$ spectra could be useful in the global analysis of $D$-meson FFs. 

\begin{figure}[t!]
    \centering
    \includegraphics[width=0.49\linewidth]{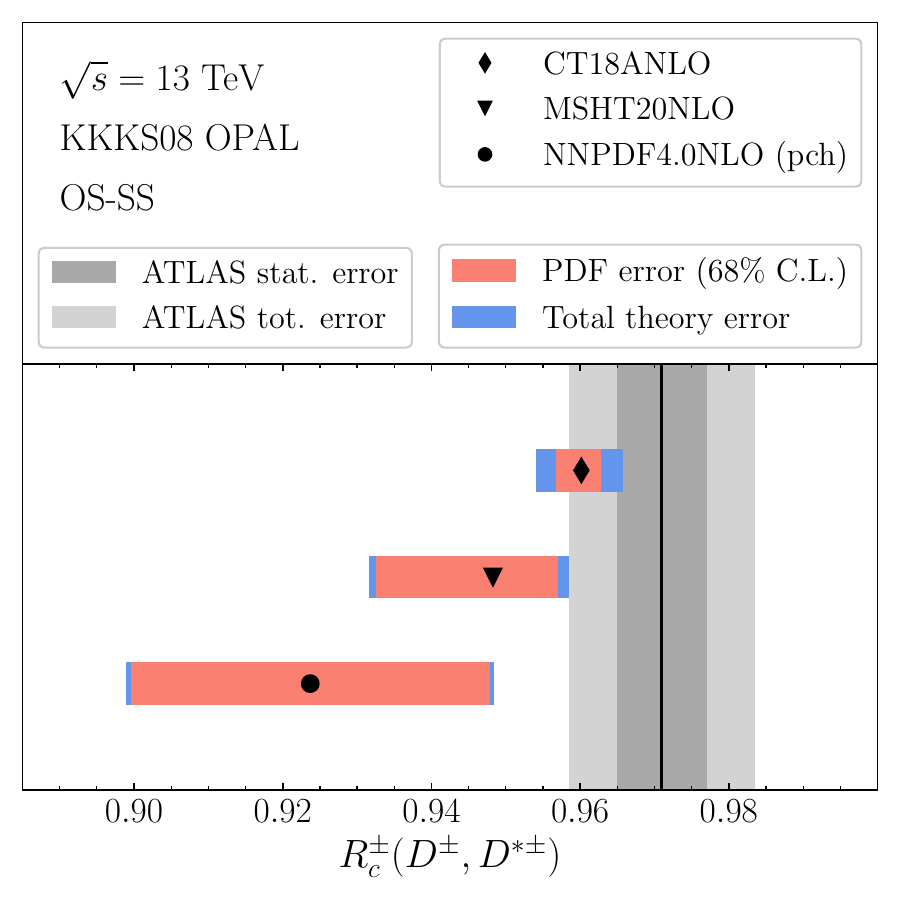}
    \includegraphics[width=0.49\linewidth]{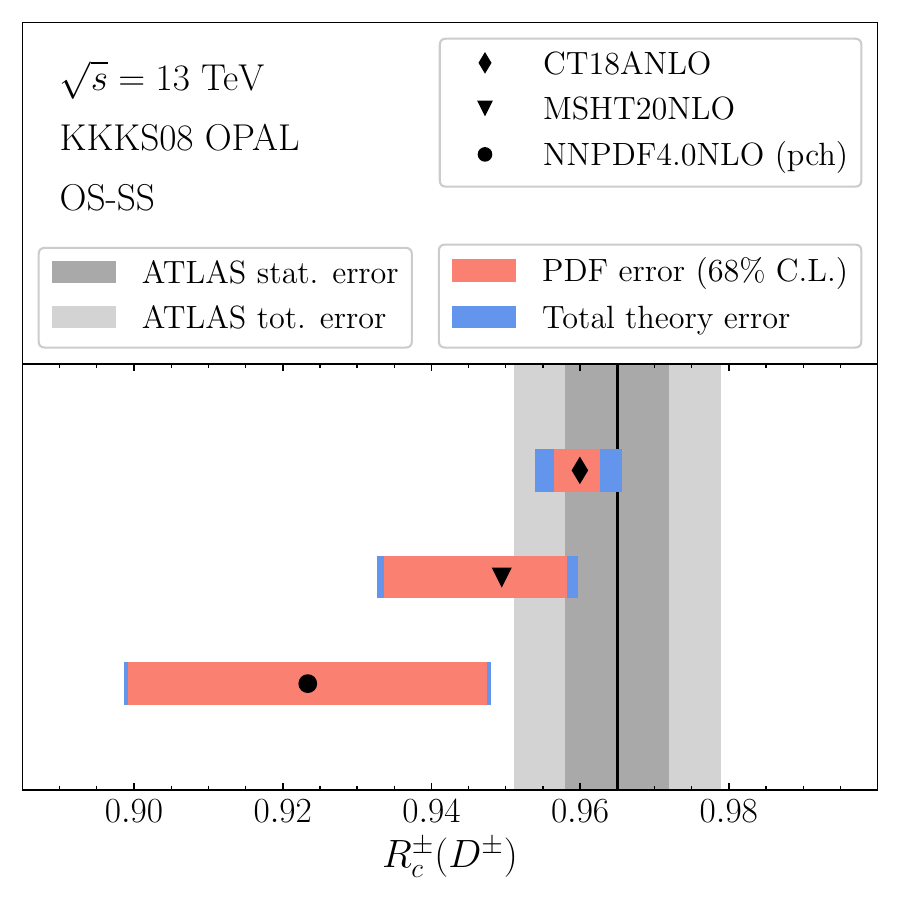}
    \caption{The ratios $R_c^\pm (D^\pm, D^{*\pm})$ and $R_c^\pm(D^\pm)$ between integrated cross sections calculated at NLO compared with the ATLAS data \cite{measurementforcomparison}.}
    \label{fig: Rcpm}
\end{figure}

\begin{figure}[htb!]
    \centering
    \begin{subfigure}{0.49\linewidth}
        \centering
        \includegraphics[width=\linewidth]{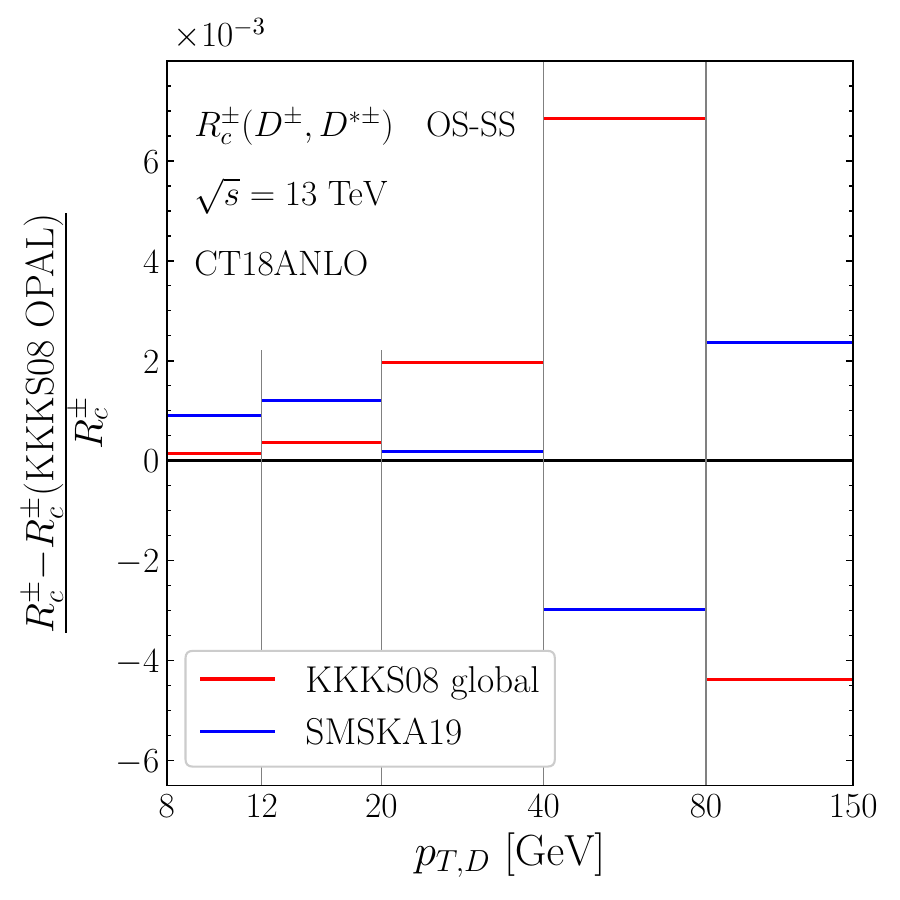}
        \caption{}
        \label{fig: Rcpm opal vs global}
    \end{subfigure}
    \begin{subfigure}{0.49\linewidth}
        \centering
        \includegraphics[width=\linewidth]{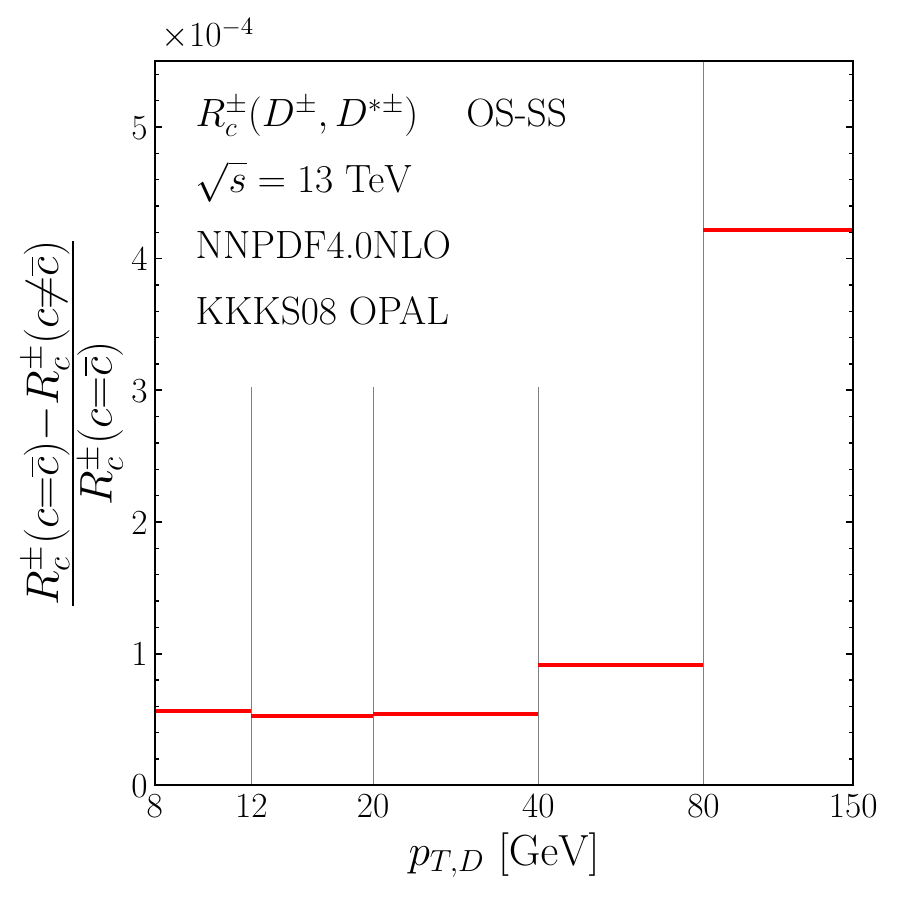}
        \caption{}
        \label{fig: c neq cbar}
    \end{subfigure}
    \caption{a) The dependence of $R_c^\pm(D^\pm, D^{*\pm})$ on the choice of the FF fit. b) The dependence of $R_c^\pm(D^\pm, D^{*\pm})$ on charm asymmetry, estimated by comparison of the PDF sets NNPDF40\_nlo\_pch\_as\_01180 ($c = \Bar{c}$) and NNPDF40\_nlo\_as\_01180 ($c \neq \Bar{c}$).}
\end{figure}

To suppress the dependence on FFs discussed above and thereby isolate the sensitivity on PDFs, we consider ratios of charge-conjugate final states, 
\begin{equation}
    \label{eq: Rcpm}
    R_c^{\pm} (h^\pm_1, \dots, h^\pm_n) = \frac{\sum_{i = 1}^n \sigma(W^++h_i^-)}{\sum_{i = 1}^n \sigma(W^-+h_i^+)} \,.
\end{equation}
In these ratios the FF dependence (see Figure~\ref{fig: Rcpm opal vs global} ahead) and also the sensitivity to the scale choices turn out to cancel almost completely. On top of this, several systematic experimental uncertainties, like the one from the luminosity, cancel between the numerator and denominator. Here, we will consider   
$R_c^\pm(D^\pm, D^{*\pm})$ and $R_c^\pm(D^\pm)$ to also study the mutual consistency between different charmed mesons. 

Figure \ref{fig: Rcpm} shows the values of the two production ratios for the three PDF sets and how they compare with the ATLAS values \cite{measurementforcomparison} in the case of fully integrated cross sections. The PDF uncertainties of $R_c^\pm$ vary quite significantly between the three considered sets of PDFs. Importantly, CT18A agrees reasonably with the measurement, especially so in the case of $R_c^\pm(D^\pm)$, while also having the smallest PDF uncertainty by far. As will be further discussed below, the smallness of the CT18A uncertainty originates from the condition $s=\overline{s}$ imposed by the CT18A fit. The other two sets of PDFs, MSHT20 and NNPDF4.0 do not impose such a constraint which already indicates that it is the not-so-well-constrained $s$ vs. $\overline{s}$ asymmetry which makes the big difference in the sizes of the PDF uncertainties. 

\begin{figure}[htb!]
    \centering
    \includegraphics[width=0.49\linewidth]{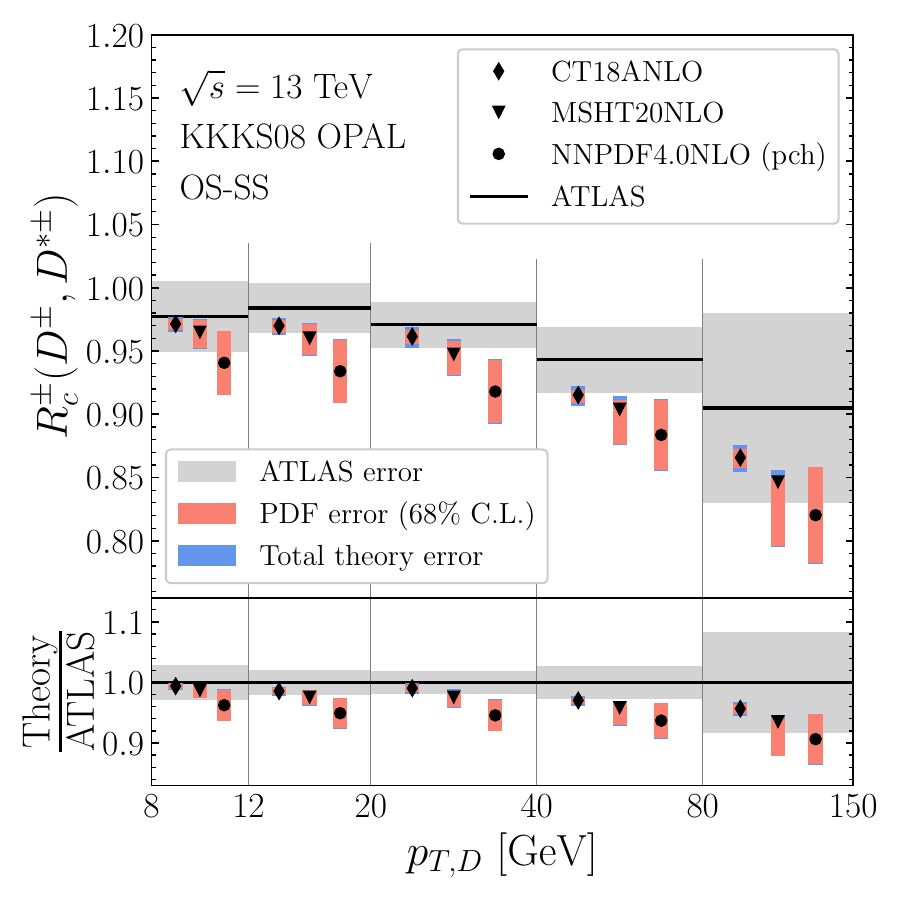}
    \includegraphics[width=0.49\linewidth]{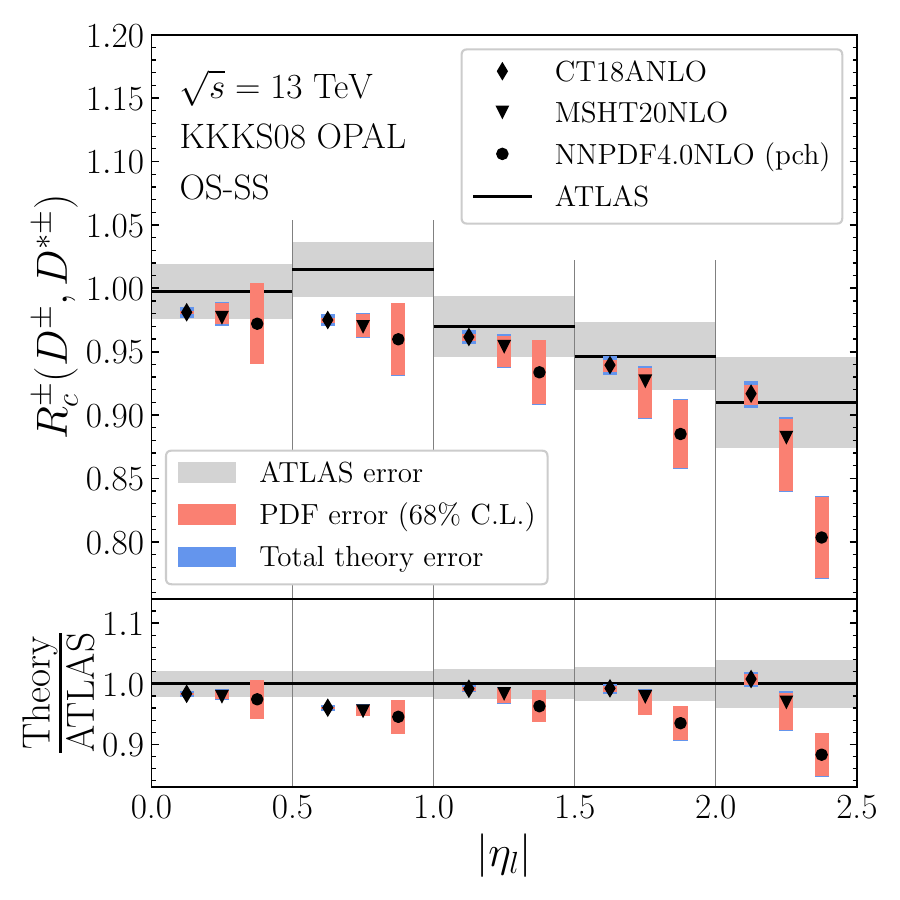}
    \includegraphics[width=0.49\linewidth]{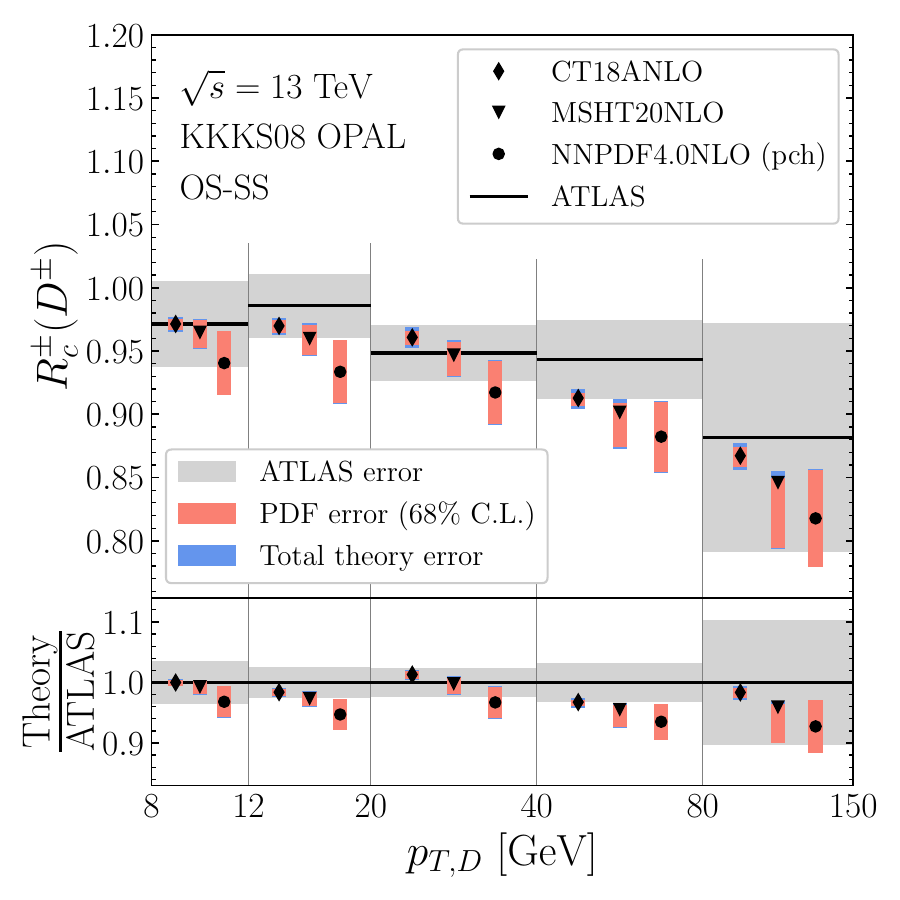}
    \includegraphics[width=0.49\linewidth]{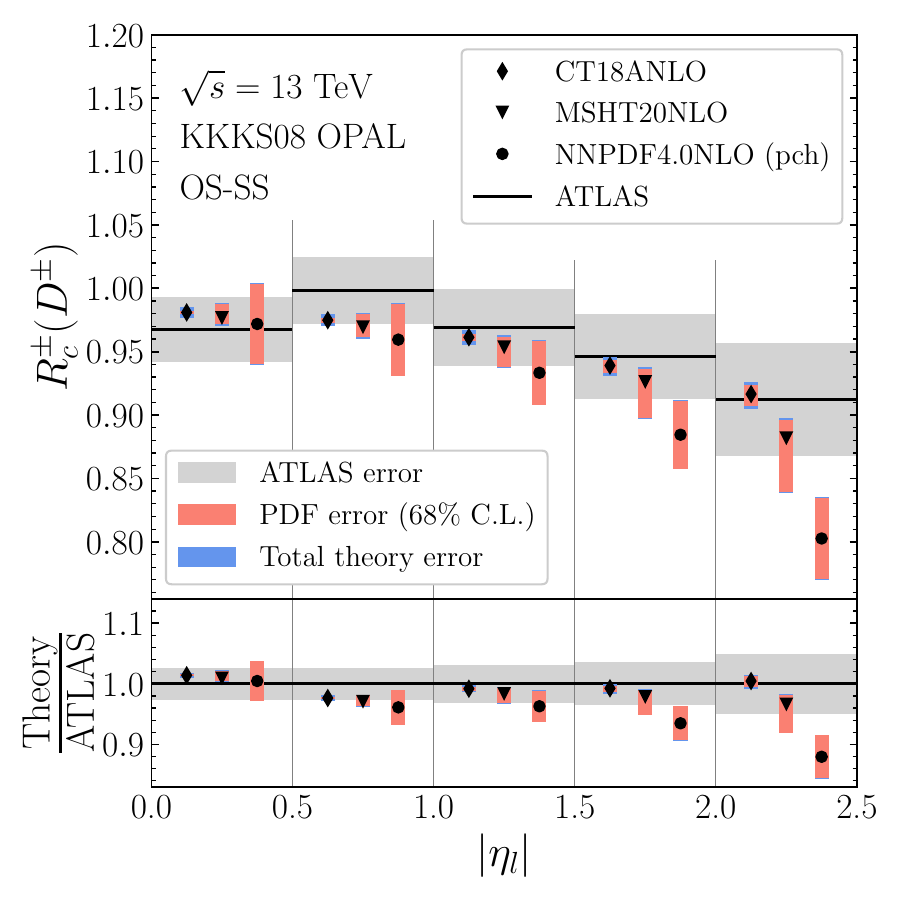}
    \caption{Production ratio $R_c^\pm$ as a function of $p_{T, D}$ and $|\eta_l|$ at NLO compared with the ATLAS data \cite{measurementforcomparison}. The upper panels include both $D^\pm$ and $D^{*\pm}$ final states, while the lower panels include only $D^\pm$.}
    \label{fig: reweighting data}
\end{figure}

\begin{table}[htb!]
\def\arraystretch{1.5}
\centering
  \caption{$\chi^2$- and $p$-values corresponding to the sum of $p_{T, D}$- and $|\eta_l|$-dependent $R_c^\pm$.}
  \begin{tabular}{ c | c | c | c | c }
      \hline
       \multirow{2}{*}{PDF set} & \multicolumn{2}{c}{$R_c^\pm (D^\pm, D^{*\pm})$} & \multicolumn{2}{|c}{$R_c^\pm (D^\pm)$} \\
        \cline{2-5}
        & $\chi^2 / \text{$N_{\rm data}$}$ & $p$-value [\%] & $\chi^2 / \text{$N_{\rm data}$}$ & $p$-value [\%]\\
        \hline
        CT18ANLO & 0.7 & 75.5 & 0.3 & 98.5 \\
        \hline
        MSHT20nlo\_as118 & 1.3 & 21.7 & 0.5 & 86.3 \\
        \hline
        NNPDF40\_nlo\_pch\_as\_01180 & 4.8 &  $5.8 \times 10^{-5}$ & 2.5 & 0.6 \\
        \hline
    \end{tabular}
    \label{tab: chi squared and p-values}
\end{table}

In Figure~\ref{fig: reweighting data} we consider the production ratios of Eq.~(\ref{eq: Rcpm}) between differential cross sections as a function of $p_{T, D}$ and $\eta_l$. We note that these observables were not explicitly considered in the original ATLAS publication \cite{measurementforcomparison} but we have formed the experimental values from the measured cross sections propagating the correlated systematic uncertainties in a standard way as done e.g. in Ref.~\cite{Eskola:2022rlm}. Also here, the NLO predictions tend to fall somewhat below the measured ratios with no significant dependence on the $D$-meson species. The overall hierarchy between the three considered sets of PDFs is also the same i.e. CT18A is on average closest to the experimental data while NNPDF4.0 disagrees the most. Also here, the CT18A PDF uncertainties remain always clearly smaller than those of MSHT20 or NNPDF4.0 due to the $s=\bar{s}$ constraint imposed in the fit. 

Table~\ref{tab: chi squared and p-values} quantifies the level of agreement seen in Figure~\ref{fig: reweighting data} in terms of $\chi^2$ and $p$ values which account for the correlated uncertainties. The numbers in the table confirm what the Figure showed: While CT18A yields a $\chi^2/N_{\rm data}$ value which always remains below unity, the corresponding values for NNPDF4.0 are much larger than unity indicating a disagreement. The MSHT20 values are somewhat larger than those of CT18A, but in statistical sense indicate an agreement with the data.

\begin{figure}[htb!]
    \centering
    \begin{subfigure}{0.49\linewidth}
        \centering
        \includegraphics[width=\linewidth]{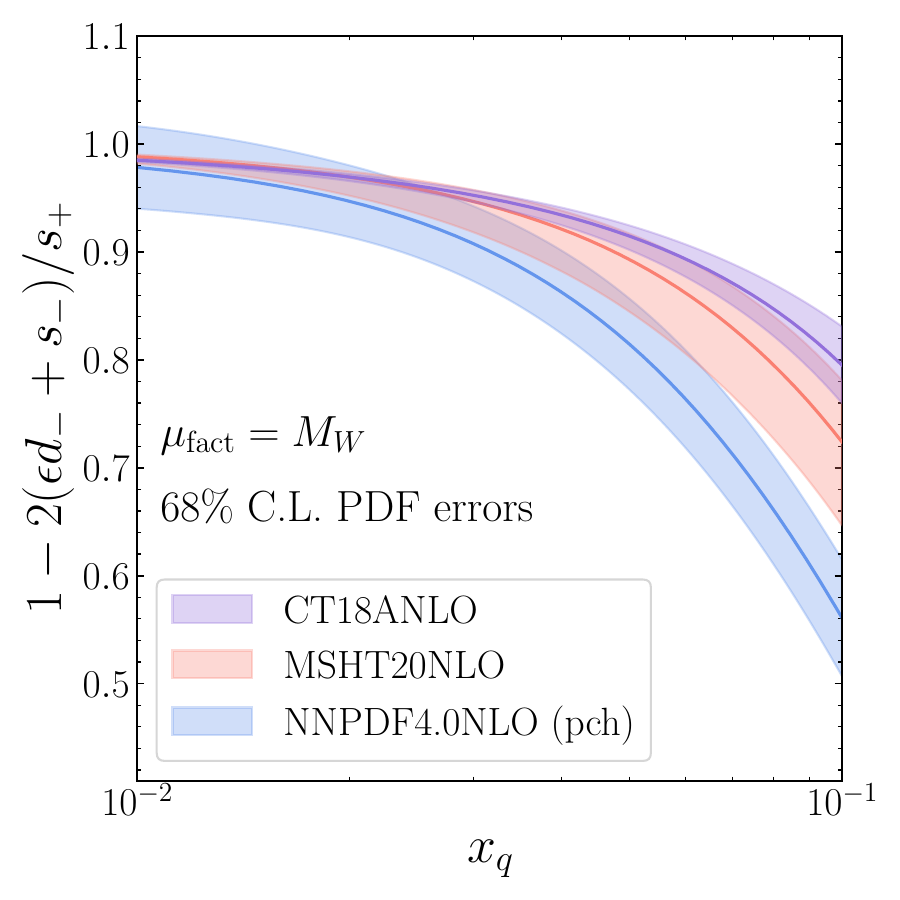}
        \caption{}
        \label{fig: Rcpm approximation}
    \end{subfigure}
    \begin{subfigure}{0.49\linewidth}
        \centering
        \includegraphics[width=\linewidth]{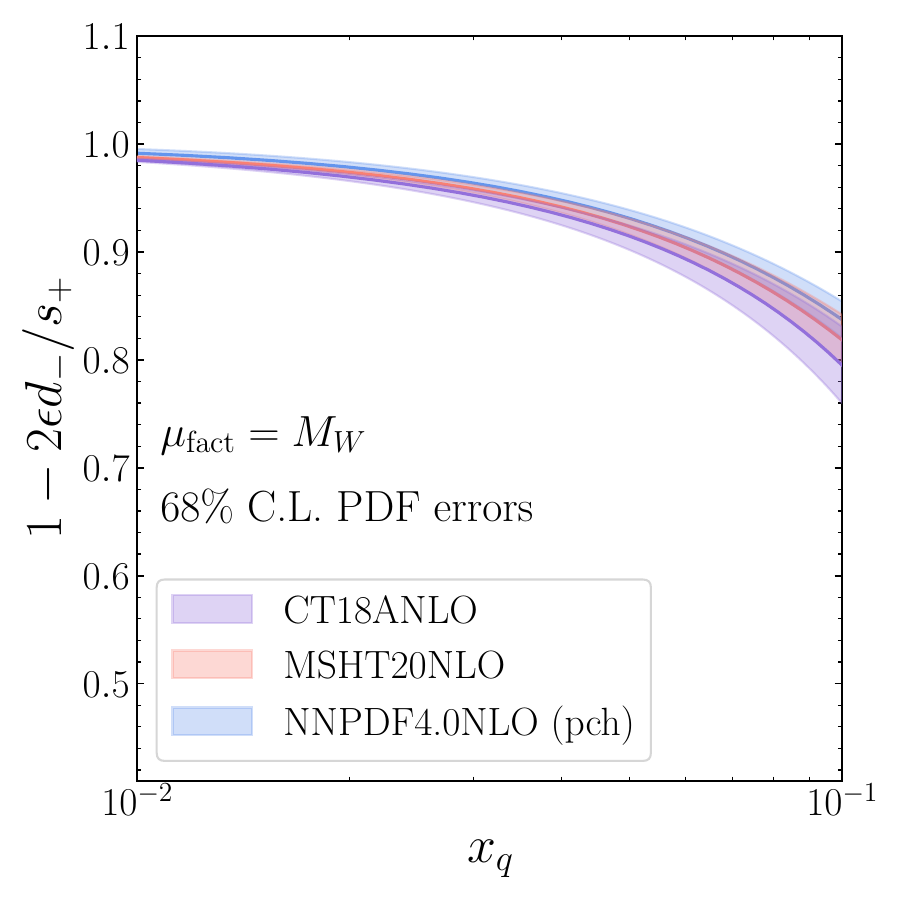}
        \caption{}
        \label{fig: Rcpm approximation s=sbar}
    \end{subfigure}
    \caption{a) Combination of PDFs in Eq.~(\ref{eq:pcomb}) for CT18ANLO, MSHT20NLO and NNPDF4.0NLO (pch).
    b) As a), but setting $s_- = 0$.
    }
    \label{fig:placeholder}
\end{figure}

The spread between the predictions of the three PDF sets in Figure~\ref{fig: reweighting data} grow larger as either $p_{T, D}$ or $|\eta_l|$ increase which suggestes that the differences in predictions should originate from the behaviour of PDFs towards larger values of $x$. To this end, we consider the cross sections to be dominated by the quark-gluon partonic processes in which the gluon comes from smaller $x$ (the gluon PDF grows rapidly towards small $x$) while the quark comes from larger $x$. By this logic
the production ratio $R_c^\pm$ should be sensitive to the following ratio of PDFs,
\begin{equation}
    \frac{|V_{cd}|^2\Bar{d} + |V_{cs}|^2\Bar{s}}{|V_{cd}|^2d + |V_{cs}|^2s} \approx 1 - 2\frac{\epsilon d_- + s_-}{s_+},
    \label{eq:pcomb}
\end{equation}
where
\begin{equation}
    \epsilon \equiv \frac{|V_{cd}|^2}{|V_{cs}|^2} \approx 0.05\,, \quad s_- \equiv s - \Bar{s} \,, \quad \quad s_+ \equiv s + \Bar{s}.
\end{equation}
The latter approximative equality comes from expanding in $\epsilon$ and $s_-/s_+$ to first order and dropping terms proportional to $\epsilon s_- / s_+$. The resulting quantity is plotted in Figure~\ref{fig: Rcpm approximation} for the three different PDF sets as a function of $x$. The resulting hierarchy is evidently reminicent of what was seen in Figure~\ref{fig: reweighting data} as $p_{T, D}$ or $|\eta_l|$ grow. The differences in Figure~\ref{fig: Rcpm approximation} can be mostly explained by the presence of non-zero $s_-$ in MSHT20 and NNPDF4.0. This is illustrated in Figure \ref{fig: Rcpm approximation s=sbar} in which the $s_-$ term has been dropped leading to a good agreement between the three considered sets of PDFs. This consolidates the idea that it is mostly the strangeness asymmetry that causes the differences seen in Figure~\ref{fig: reweighting data} between the considered PDF sets and that the charge ratios could thereby be a powerful constraint on the strangeness asymmetry.

\begin{figure}[t!]
    \centering
    \includegraphics[width=0.49\linewidth]{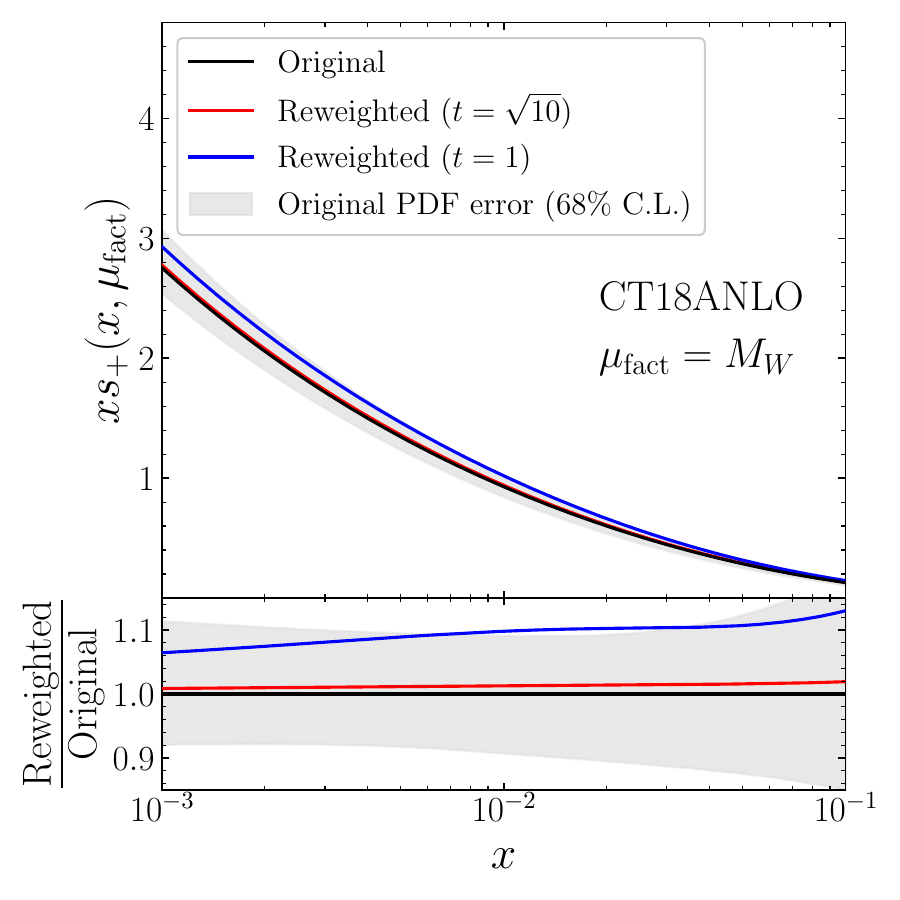}
    \includegraphics[width=0.49\linewidth]{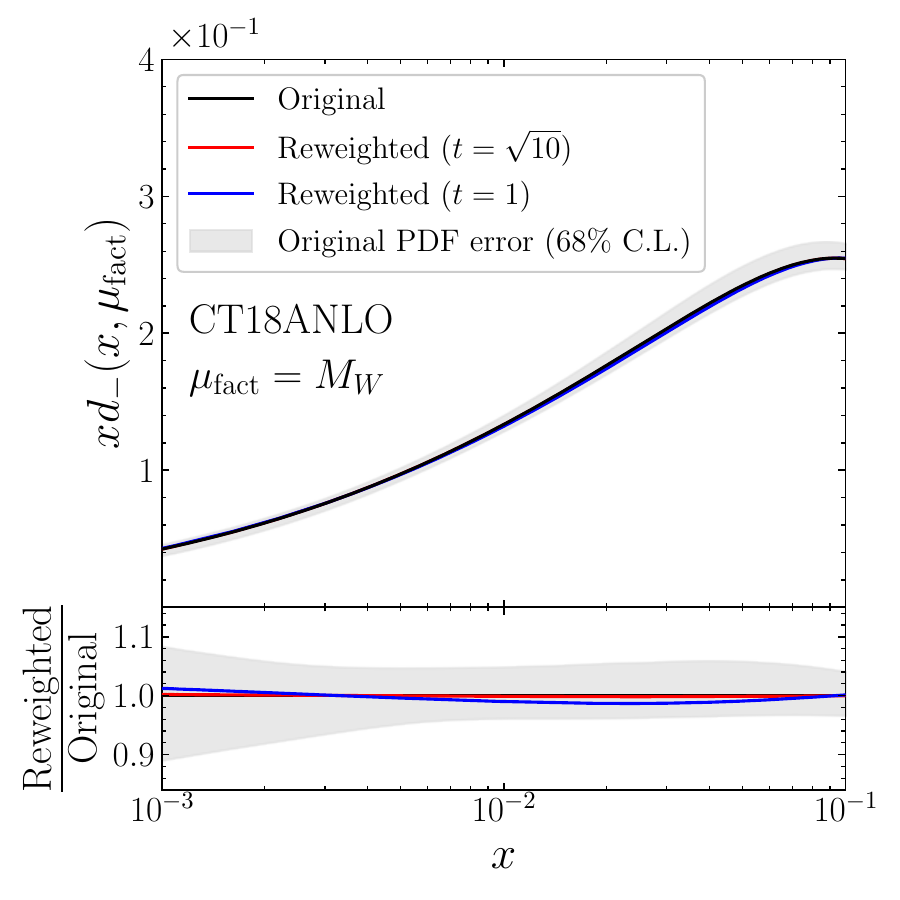}
    \includegraphics[width=0.49\linewidth]{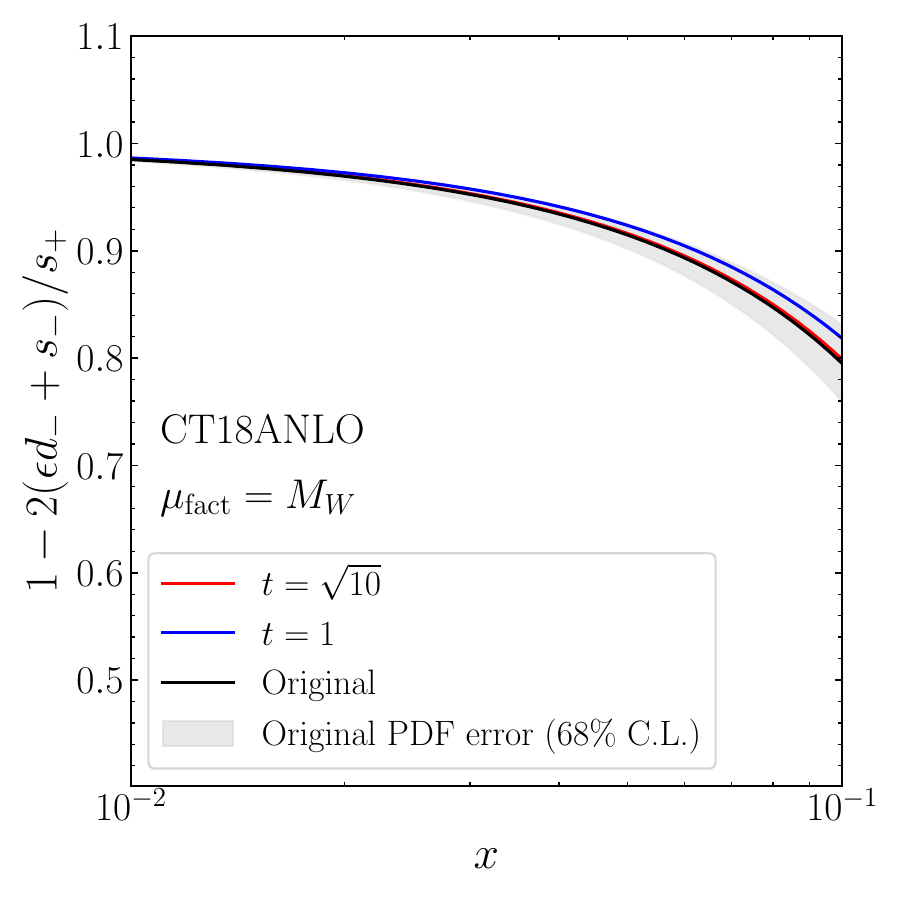}
    \caption{Reweighting CT18ANLO with the $\eta_l$- and $p_{T, D}$-dependent $R_c^\pm(D^\pm, D^{*\pm})$.}
    \label{fig: reweighted CT18ANLO}
\end{figure}


\begin{figure}[htb!]
    \centering
    \includegraphics[width=0.49\linewidth]{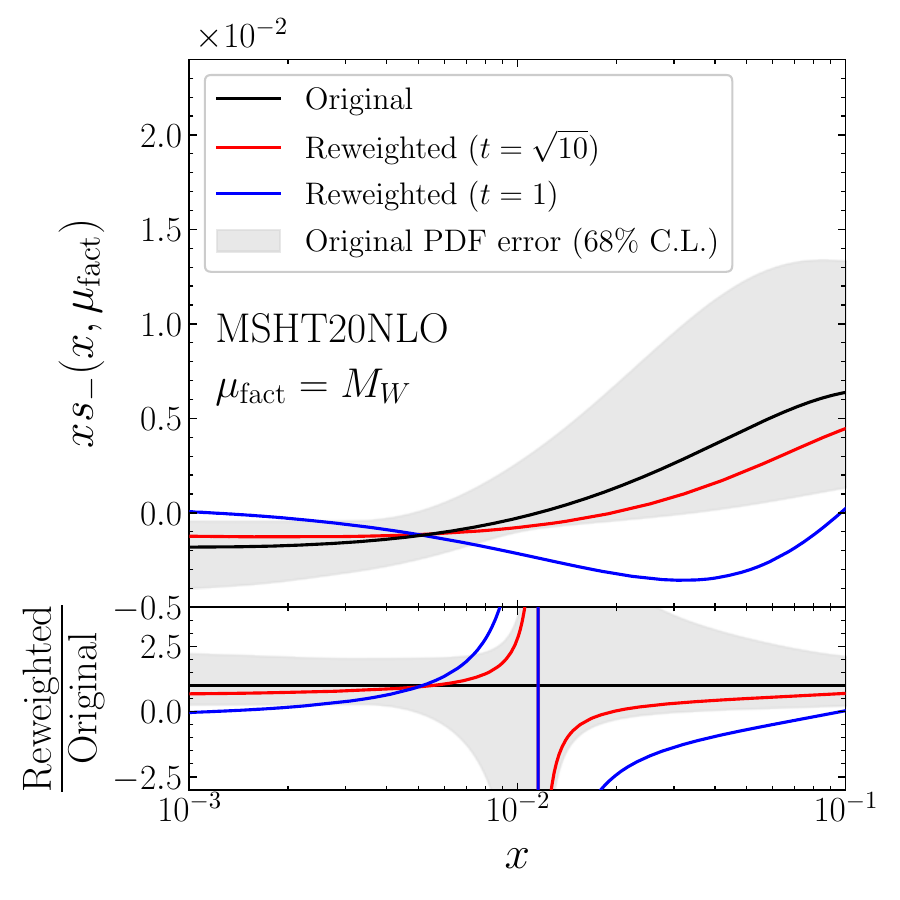}
    \includegraphics[width=0.49\linewidth]{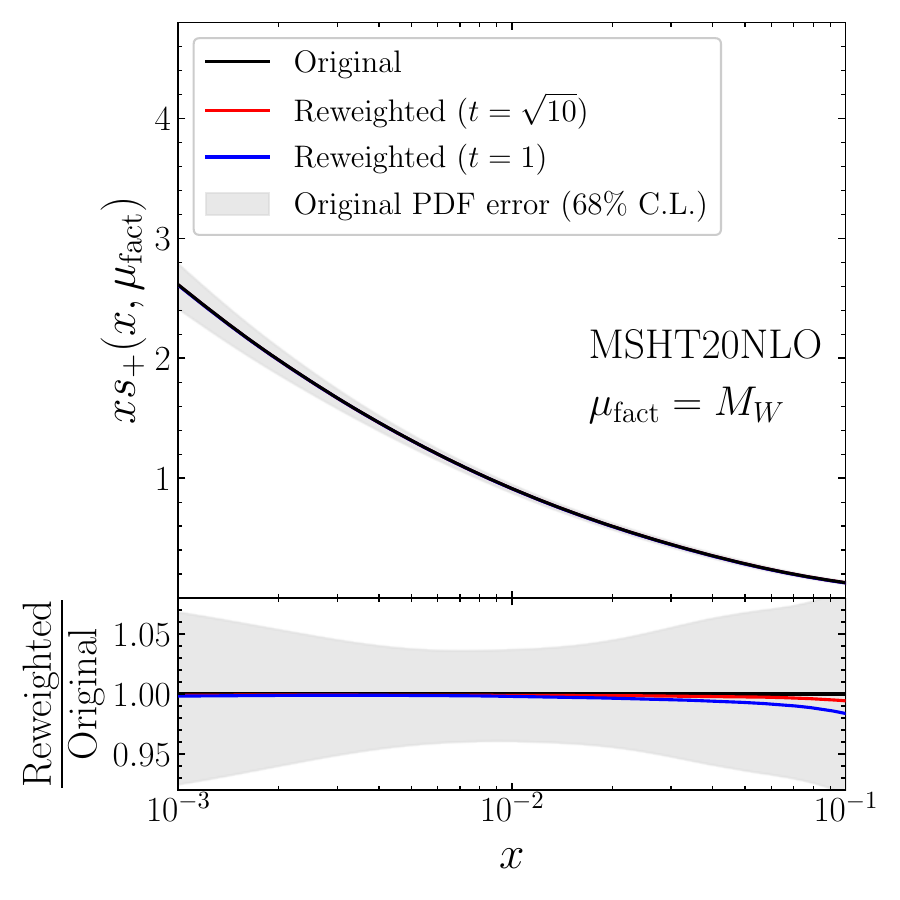}
    \includegraphics[width=0.49\linewidth]{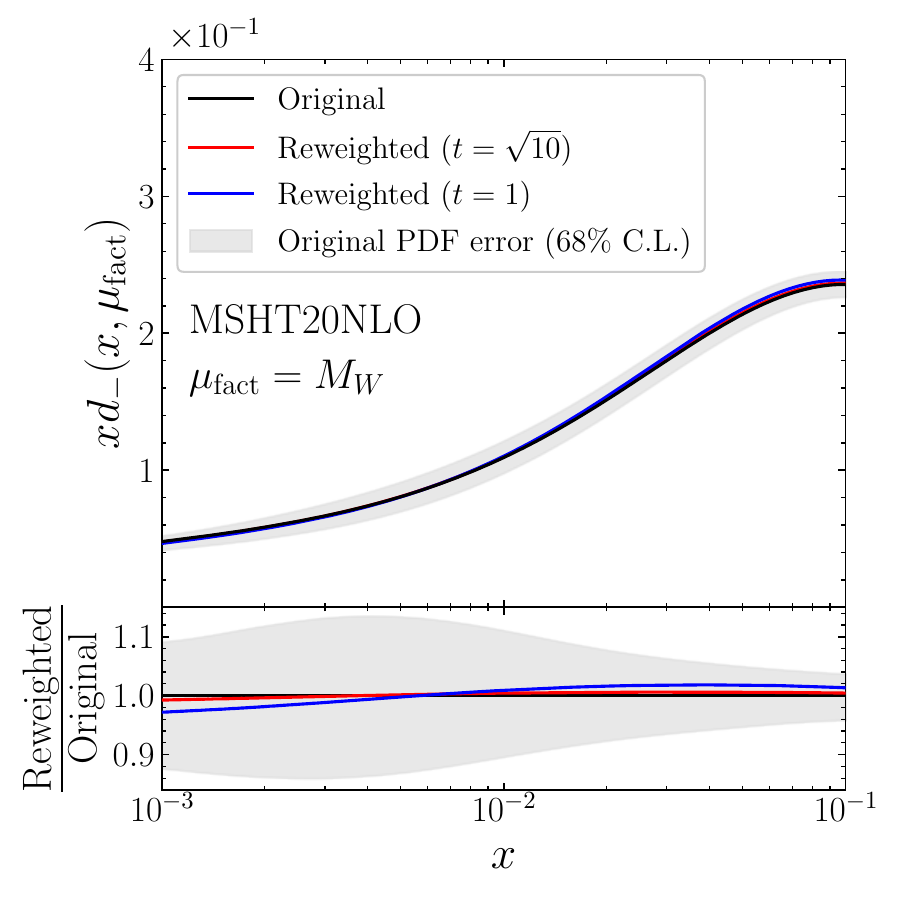}
    \includegraphics[width=0.49\linewidth]{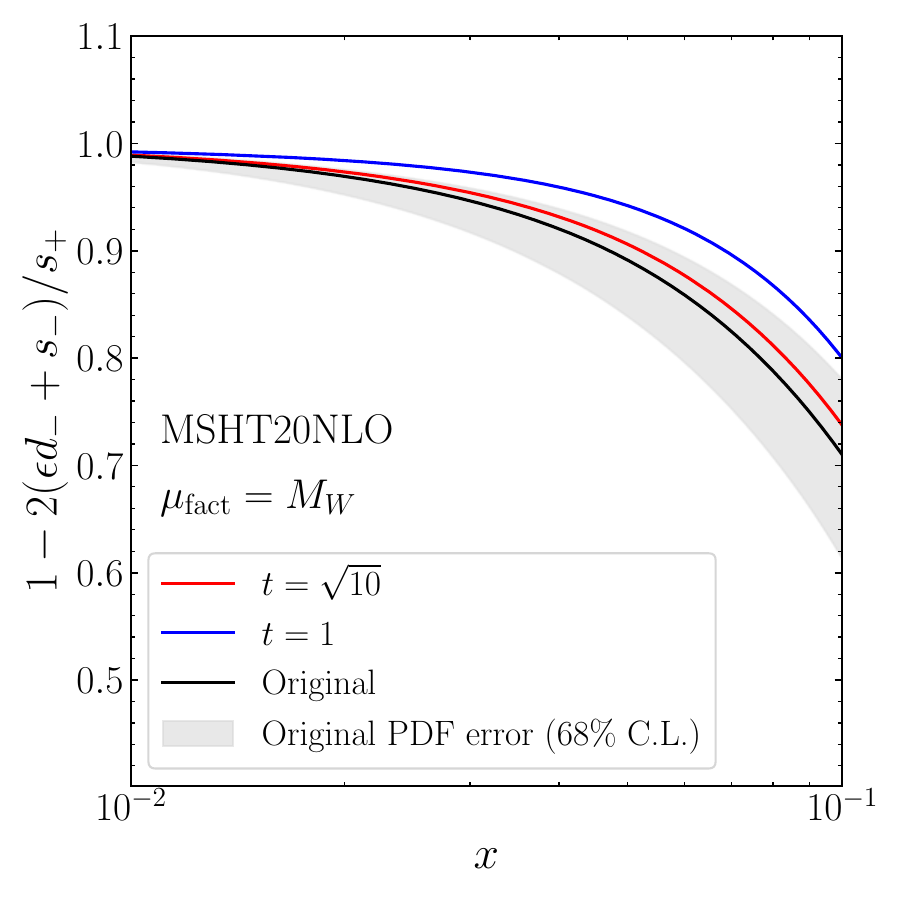}
    \caption{Reweighting MSHT20NLO with the $\eta_l$- and $p_{T, D}$-dependent $R_c^\pm(D^\pm, D^{*\pm})$.
    }
    \label{fig: reweighted MSHT20NLO}
\end{figure}


\begin{figure}[htb!]
    \centering
    \includegraphics[width=0.49\linewidth]{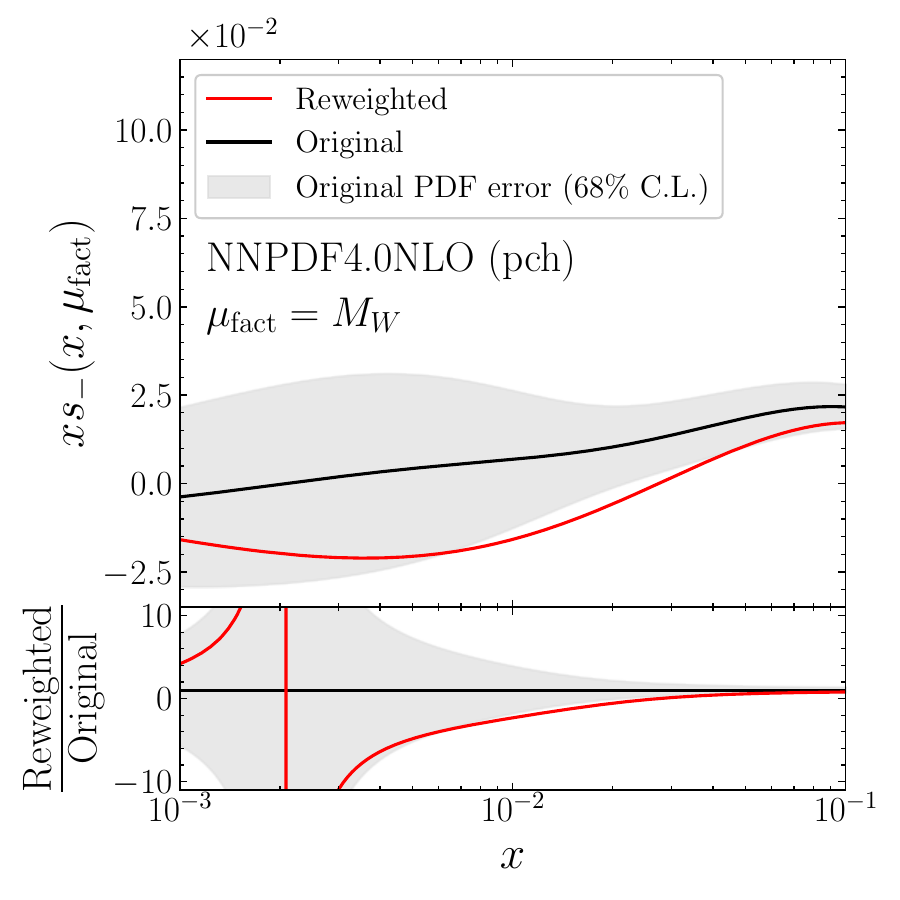}
    \includegraphics[width=0.49\linewidth]{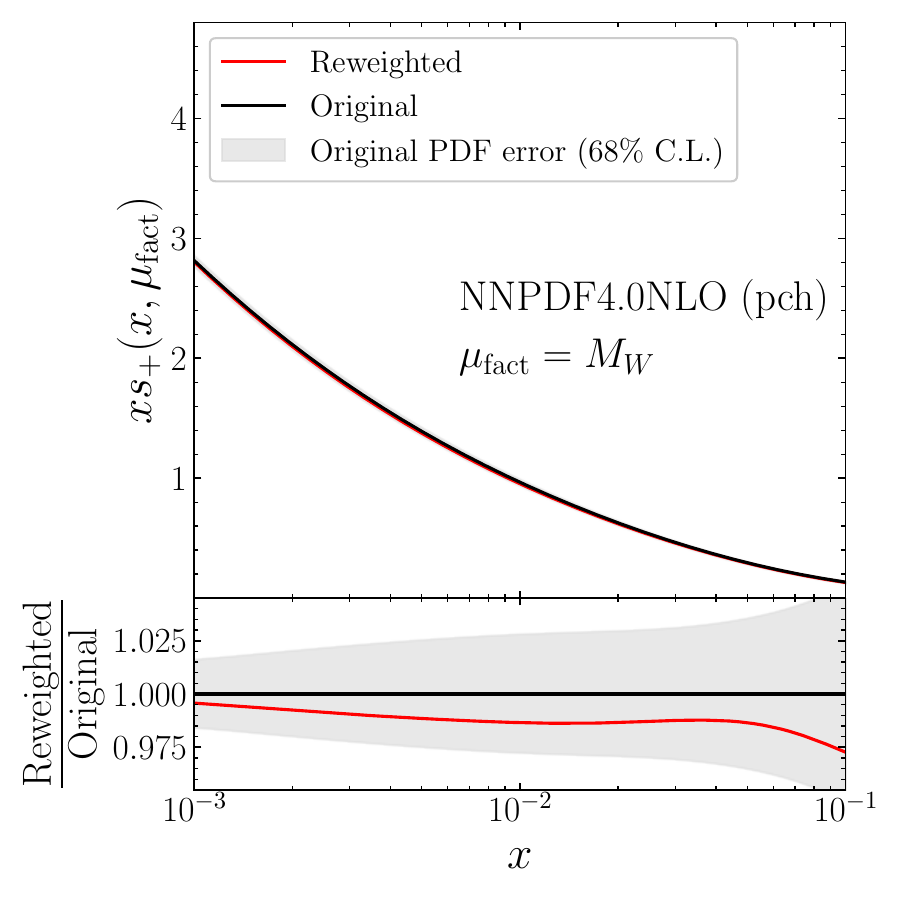}
    \includegraphics[width=0.49\linewidth]{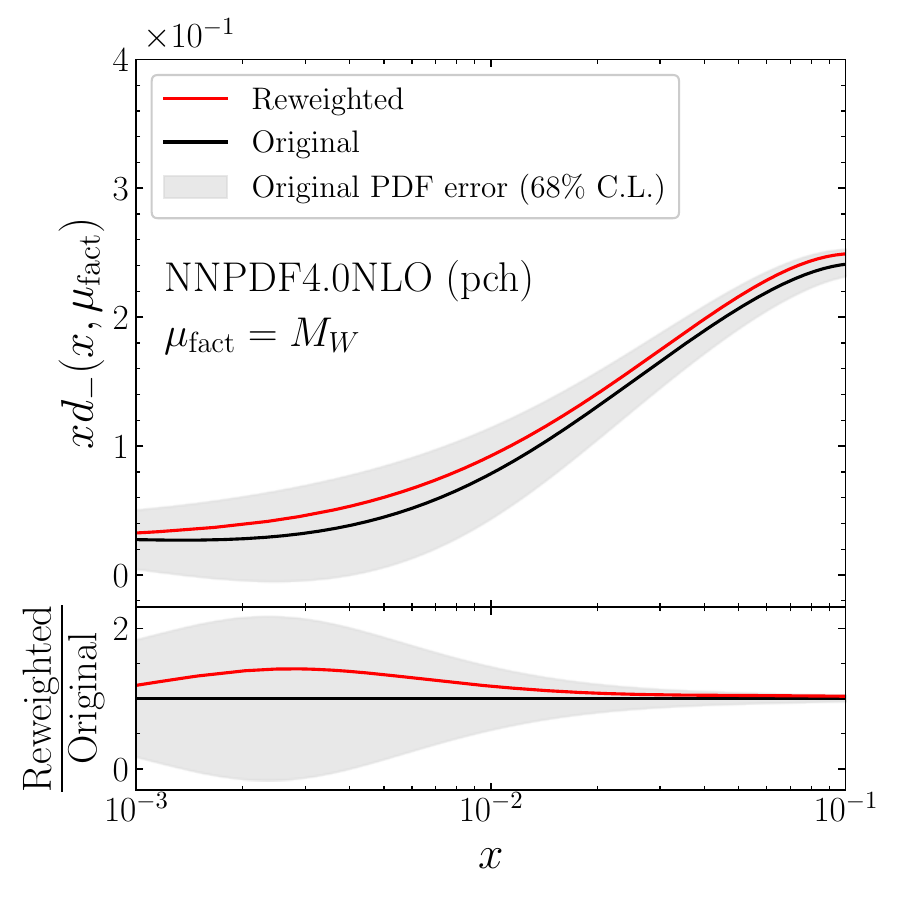}
    \includegraphics[width=0.49\linewidth]{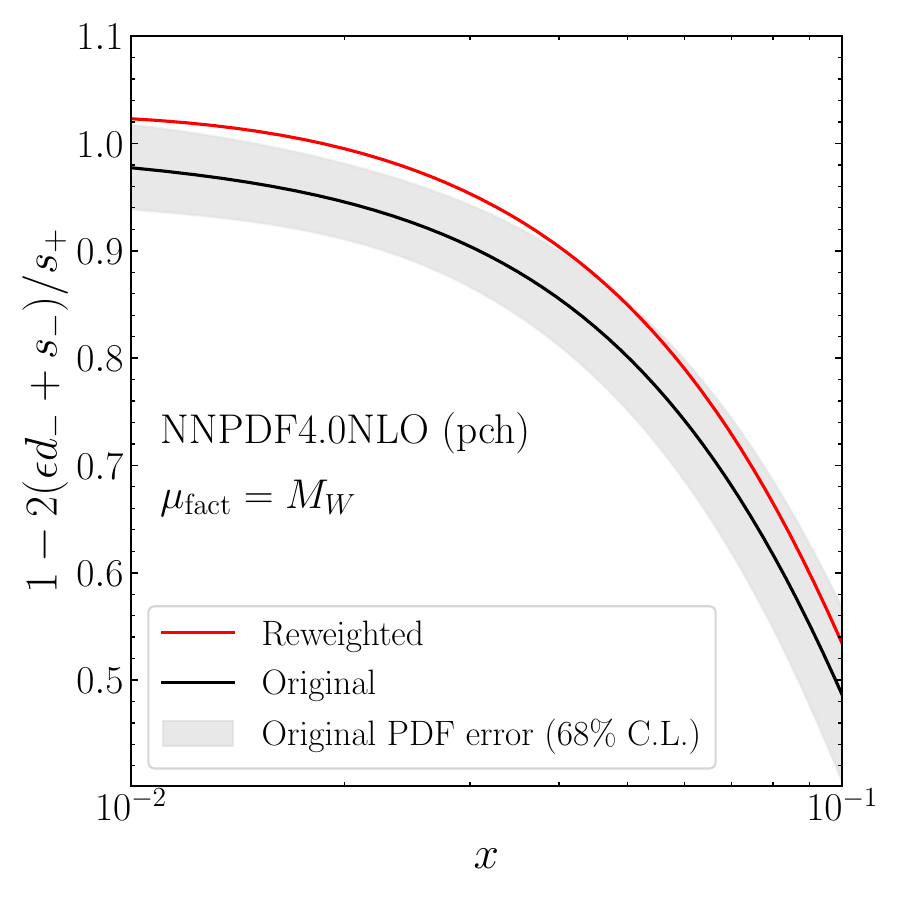}
    \caption{Reweighting NNPDF4.0NLO (pch) with the $\eta_l$- and $p_{T, D}$-dependent $R_c^\pm(D^\pm, D^{*\pm})$.
    }
    \label{fig: reweighted NNPDF4.0NLO(pch)}
\end{figure}

\begin{figure}[htb!]
    \centering
    \includegraphics[width=0.49\linewidth]{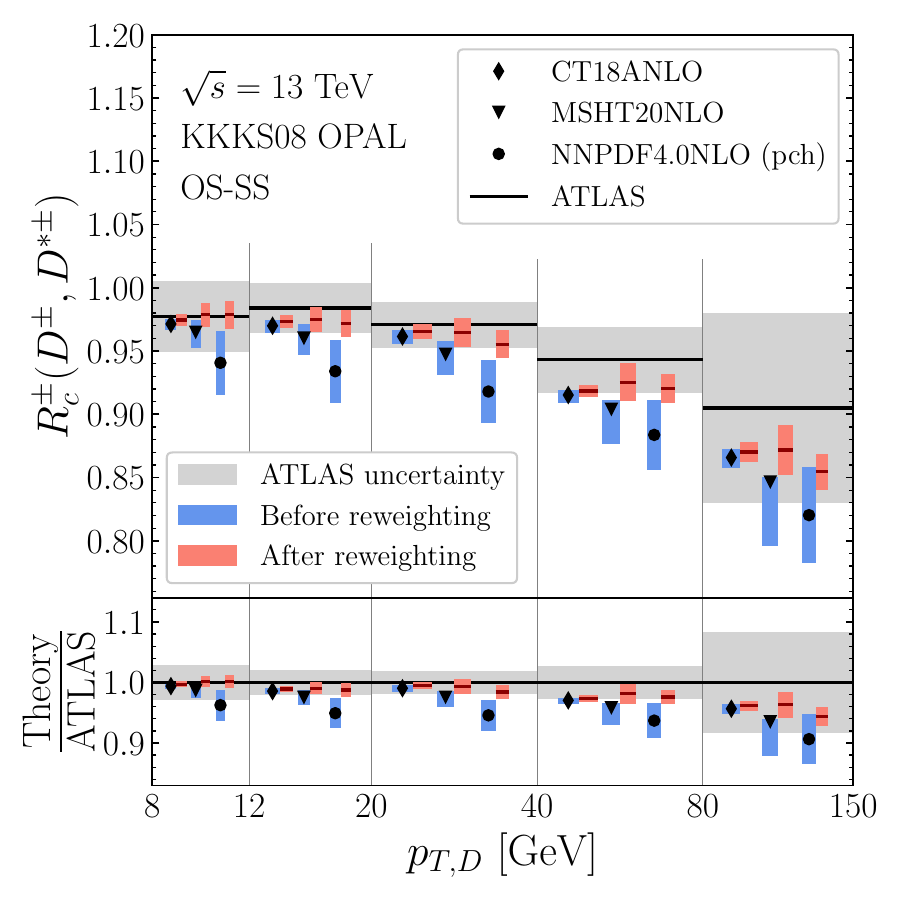}
    \includegraphics[width=0.49\linewidth]{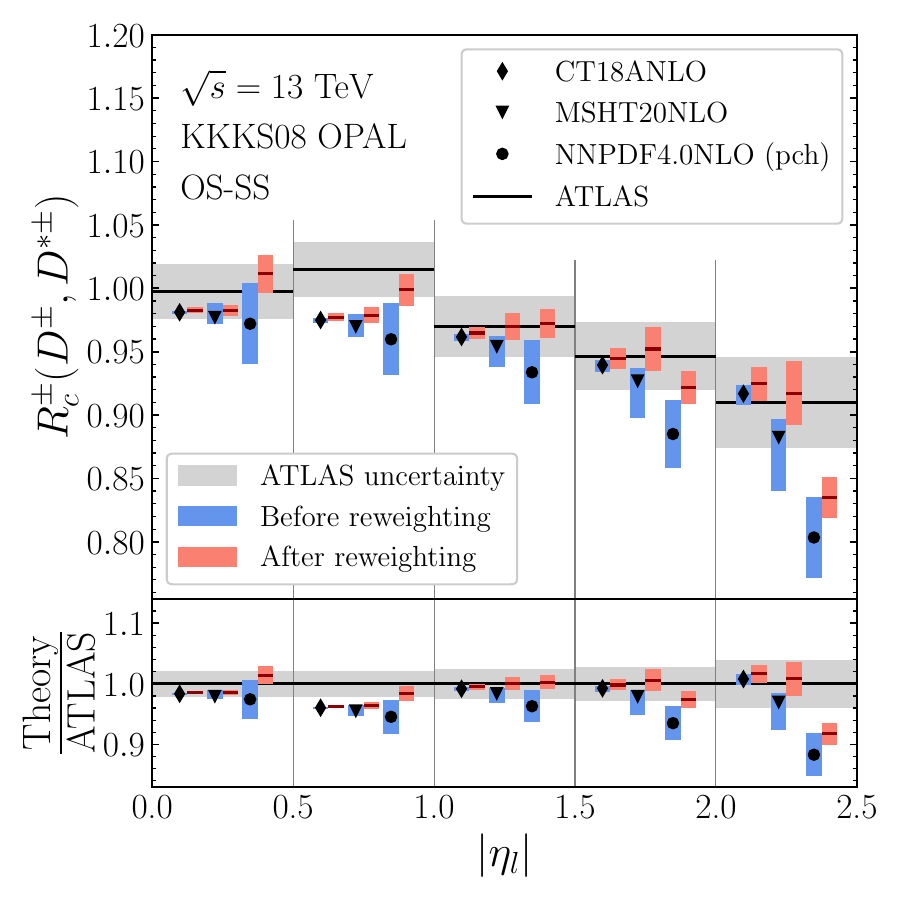}
    \caption{Effects of reweighting on the bin-integrated $R_c^\pm(D^\pm, D^{*\pm})$. Tolerance pa\-ra\-me\-ter $t = 1$ is used for the Hessian PDF sets CT18A and MSHT20.}
    \label{fig: Rcpm reweighted}
\end{figure}

Because the theory predictions for $R_c^\pm$ fall below the data, a better agreement would require some combination of larger $s_+$, smaller $d_-$ and/or smaller $s_-$. To make this more concrete we use here the standard PDF reweighting/profiling methods. In the case of CT18A and MSHT20 we use linear Hessian reweighting \cite{Paukkunen_2014,Eskola:2019dui} and for NNPDF4.0 the bayesian reweighting \cite{Ball:2010gb,Ball:2011gg}. The data points included in the reweighting study here are the $R_c^\pm(D^\pm, D^{*\pm})$ data points as a function of $p_{T, D}$ and $|\eta_l|$ -- the ten data points in the first row of Figure~\ref{fig: reweighting data}. There is some statistical overlap between these two as most of the cross section which is distributed to $|\eta_l|$ bins accumulate from the first bins in $p_{T, D}$. However, we are here not after a statistically rigorous analysis but want to obtain a rough idea how the PDFs should change to better reproduce the $R_c^\pm(D^\pm, D^{*\pm})$ data. The systematic correlations are taken into account when evaluating the $\chi^2$ values required in the reweighting. The ten data points to which we reweight here would carry a rather small weight in the total $\chi^2$ budget of a true global analysis. To better see to what direction these data pull the PDFs we therefore also play around with the tolerance parameter which needs to be specified in the case of Hessian reweighting: we adopted two values $t = \sqrt{10}$ and $t = 1$, the former value being a representative of the average tolerance of the MSHT20 fit. The smaller the tolerance parameter, the more impact these 10 data points have. 

Figures \ref{fig: reweighted CT18ANLO}, \ref{fig: reweighted MSHT20NLO} and \ref{fig: reweighted NNPDF4.0NLO(pch)} show how the three PDF sets and the combination of PDFs in Eq.~(\ref{eq:pcomb}) change when reweighting is applied, confirming the conjectures we made above. In the case of CT18A, which sets $s_- = 0$, the largest pull comes to $s_+$ which grows as expected, leaving $d_-$ largely unchanged. As a result, the combination of partons in Eq.~(\ref{eq:pcomb}) increases slightly and the same happens for $R_c^\pm(D^\pm, D^{*\pm})$ shown in Figure~\ref{fig: Rcpm reweighted}, where the cases $t = 1$ are shown for the Hessian PDF sets. For MSHT20 and NNPDF4.0, the largest pull is on $s_-$ which decreases towards large $x$ to better reproduce the experimental data. Logically, the combination of partons in Eq.~(\ref{eq:pcomb}) again increase for both sets of PDFs and likewise does $R_c^\pm(D^\pm, D^{*\pm})$ shown in Figure~\ref{fig: Rcpm reweighted}. We note that the starting point here was a case in which the data points tend to lie outside the PDF error bands so that the reweighting tools are beginning to be more extrapolative and arguably lose their accuracy. In case like this, a refit of PDFs would be required for a more conclusive statement of the impact of the data. In any case, the reweighting study here quite clearly indicates that these $W$+charm data should be quite powerful in constraining the $s_+$ and $s_-$ PDFs of the proton, the case of small or zero strangeness asymmetry and increased overall strangeness $s_+$ being favored. We have checked that these conclusions do not depend on the used set of FFs. This is illustrated in Figure~\ref{fig: Rcpm opal vs global} which compares the $p_{T, D}$-dependent charge ratios with three sets of FFs which were tested already in Figure~\ref{fig: sigma pTD opal vs global} in the case of absolute cross sections. As can be seen, the dependence on FFs is only at the level of permilles at least when $D_{c \rightarrow D^+} = D_{\overline{c} \rightarrow D^-}$ is assumed as done in the available FF analyses.

In our calculation, we have discarded the set of Feynman diagrams which contain $c$ or $\Bar{c}$ in the initial state, like the ones shown in Figure \ref{fig: c/cbar initial states}. If there is no intrinsic charm contribution in protons, this class of diagrams cancels exactly in the OS-SS subtraction at NLO. This remains to be the case if intrinsic charm is introduced with the constraint $c = \Bar{c}$. Only if there is charm asymmetry, $c \neq \Bar{c}$, can this class of diagrams has a non-zero contribution. We tested the magnitude of such contributions by computing cross sections with NNPDF40\_nlo\_pch\_as\_01180 ($c = \Bar{c}$) and NNPDF40\_nlo\_as\_01180 ($c \neq \Bar{c}$). One could hope that $R_c^\pm$ would be sensitive enough to this asymmetry. 
However, this turns out not to be the case, as can be seen in Figure \ref{fig: c neq cbar}: the relative difference between the differential production ratios is of the order $10^{-4}$.

\begin{figure}[htb!]
    \centering
    \begin{subfigure}{0.49\linewidth}
        \centering
        \includegraphics[width=\linewidth]{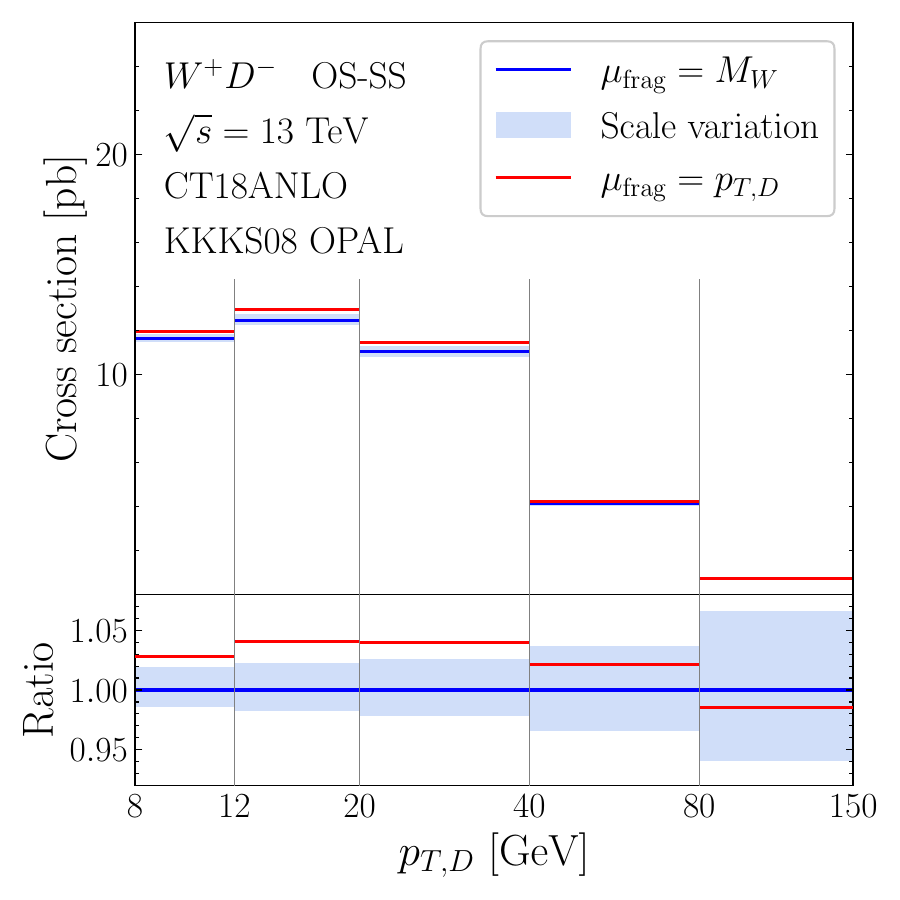}
        \caption{}
        \label{fig: pTD frag scale}
    \end{subfigure}
    \begin{subfigure}{0.49\linewidth}
        \centering
        \includegraphics[width=\linewidth]{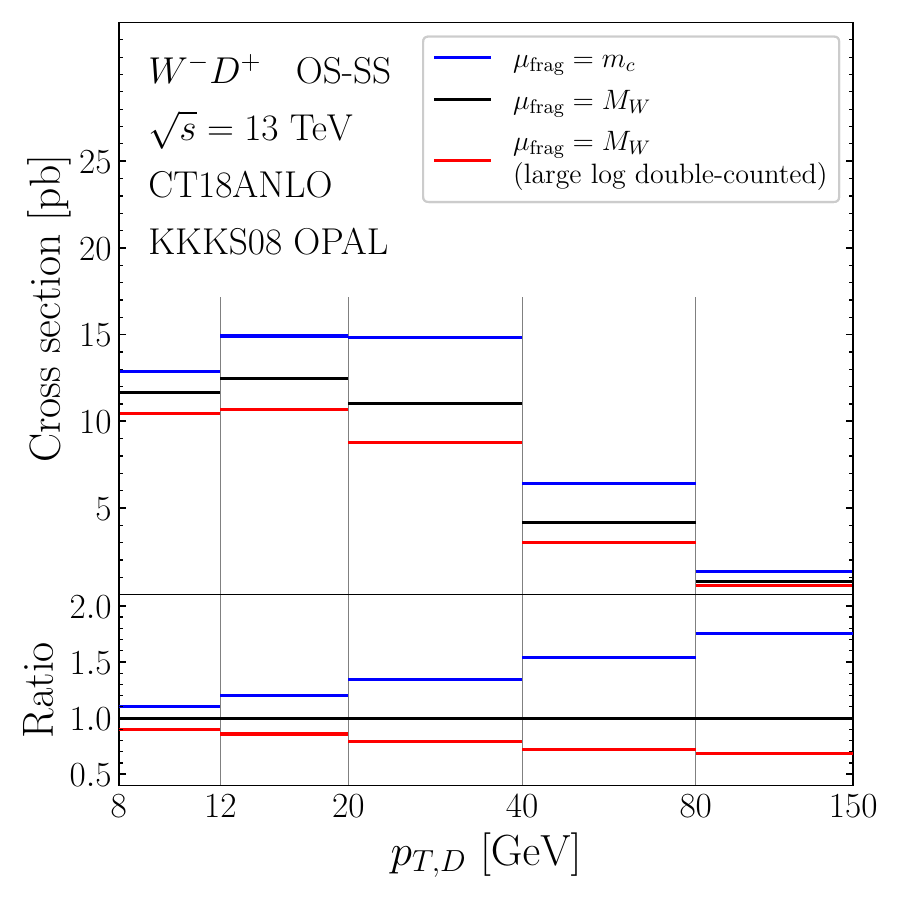}
        \caption{}
        \label{fig: effect of subtraction}
    \end{subfigure}
    
    \caption{a) Comparison between a constant and a $p_{T, D}$-dependent fragmentation scales. b) 
      The effects of FF DGLAP evolution and the subtraction term in the $p_{T, D}$ bin-integrated $W^-D^+$ cross section (see text for details).
    }
\end{figure}

Our results are calculated with a fixed fragmentation scale $\mu_\text{frag} = M_W$. Since the partonic cross sections at NLO effectively develop a $\log(p_{T, c}/m_c)$ type of logarithmic behaviour due to final state radiation, tying the fragmentation scale to the transverse momentum of the $D$ meson is also justifiable. At the lower end of the $p_{T, D}$ spectrum this naturally has some effect. 
Figure \ref{fig: pTD frag scale} shows how the $W^-D^+$ bin-integrated cross section is affected by this change in the fragmentation scale. Despite the large differences in $\mu_\text{frag}$ at low $p_{T, D}$, the values with $\mu_\text{frag} = p_{T, D}$ are almost enclosed by the scale variation around $\mu_\text{frag} = M_W$. In addition, we have checked that the differences between the two scale choices effectively vanish in the charge ratio, and therefore the choice of the fragmentation scale does not really change the implications on PDFs discussed above. 

Figure \ref{fig: effect of subtraction} quantifies the relevance of the resummation which was explained in Section \ref{sec: frag}. The blue values correspond to setting $\mu_\text{frag} = m_c$ i.e. the calculation is effectively performed in a fixed-flavour-number scheme containing the $\log(p_{T, c}/m_c)$ type of term (no resummation). Evaluating the FFs at $\mu_\text{frag} = M_W$ but leaving $\mu_\text{frag} = m_c$ in the subtraction term, the second line in Eq.~(\ref{eq:lopullinen}), gives the red values corresponding to "double counting" the first logarithmic term. Finally, when we set $\mu_\text{frag} = M_W$ in all places gives the black curves, the "correct" ones. We can understand that the resummation has a significant effect -- nearly a factor of two at large $p_{T, D}$ -- and it also changes the shape of the $p_{T, D}$ distribution in a non-trivial way. Thus, taking it consistently into account, either by the GM-VFNS framework considered here or by a parton-shower approach, is essential. 

\begin{figure}[t!]
    \centering
    \begin{subfigure}{0.49\textwidth}
        \centering
        \includegraphics[width=\linewidth]{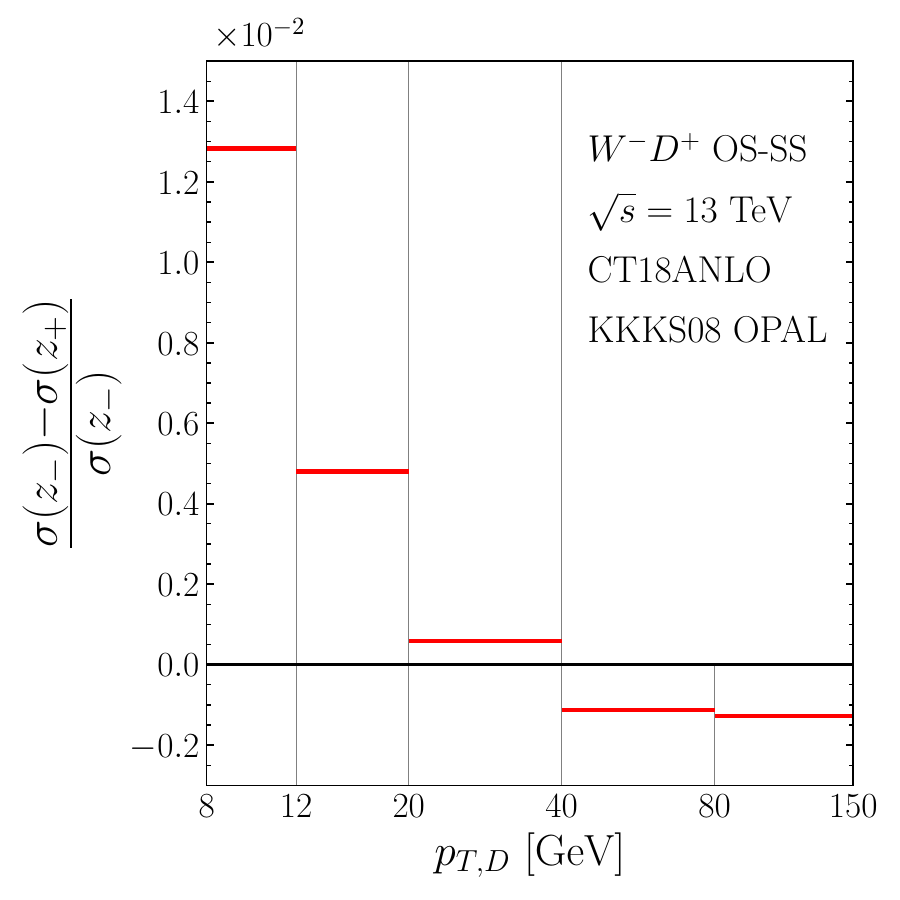}
        \caption{}
        \label{fig: zmzp}
    \end{subfigure}
    \begin{subfigure}{0.49\textwidth}
        \centering
        \includegraphics[width=\linewidth]{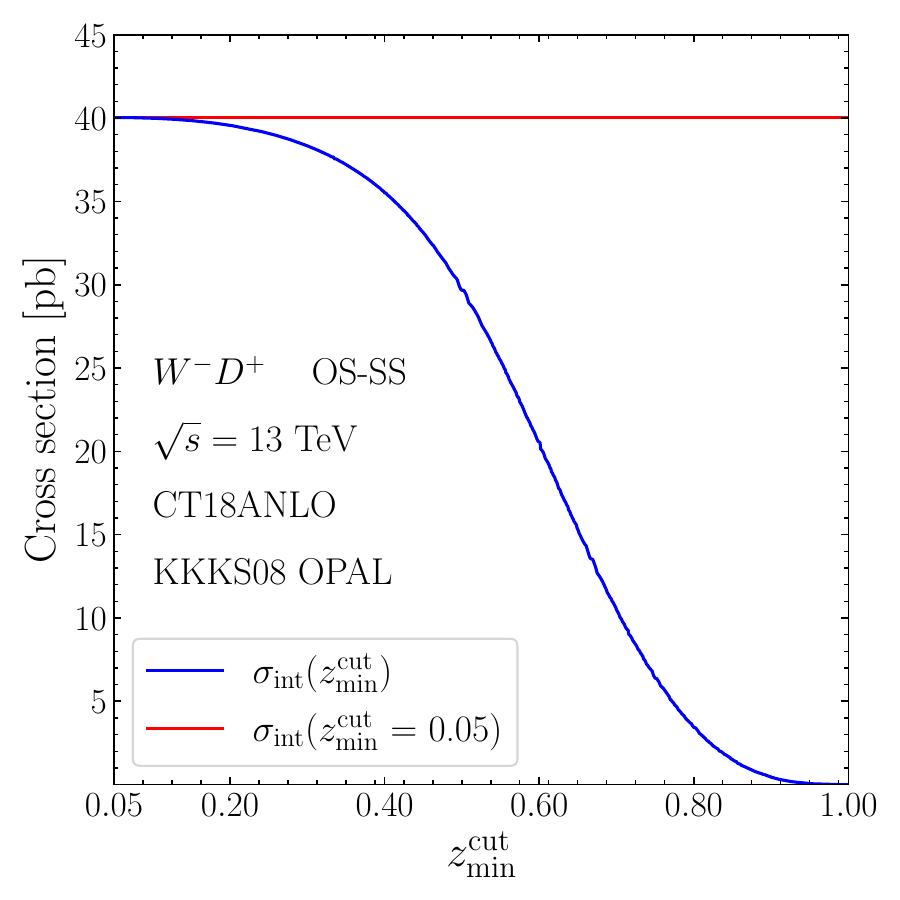}
        \caption{}
        \label{fig: zmin}
    \end{subfigure}
    \caption{a) The dependence of the $p_{T, D}$ bin-integrated $W^-D^+$ cross section on whether $z_-$ \eqref{eq: z definition} or $z_+$ \eqref{eq: z definition KKKS08} is used. The values shown here are representative of the corresponding values for the three other processes. b) The dependence of the integrated $W^-D^+$ cross section on a hard cut on the fragmentation variable $z$ from below, above which $z_\text{min}$ follows equation \eqref{eq: zmin}.}
\end{figure}

As discussed in Section \ref{sec: frag}, we chose as the fragmentation variable the ratio between the transverse momenta of the meson and the quark, $z_- = p_{T, D} / p_{T, c}$. 
Figure \ref{fig: zmzp} displays the difference between this, and the other possible choice $z_+ = E_D / E_c$ at the cross section level. As we can see, the difference is at most $\approx 1.3\%$ at the lowest $p_{T, D}$ bin and around per-mille level at high values of $p_{T, D}$. Furthermore, these differences are similar for all four processes, thus effectively canceling in the production ratios.

Another choice mentioned in Section \ref{sec: frag} was that we apply a hard cut on the fragmentation variable from below, $z_\text{min}^\text{cut}$. The KKKS08 FFs are fitted only down to $z = 0.05$, which is almost always larger than the actual kinematical limit given in equation \eqref{eq: zmin}. However, the integrands for the cross sections are highly dominated by the mid-$z$ region. This is demonstrated in Figure \ref{fig: zmin} which shows how the integrated cross section for $W^-D^+$ production (the other cases follow similar trends) accumulates as a function of $z$. In this case,
only $1\%$ of the integrated cross section comes from the range $z \in [0.05, 0.15]$. Therefore, discarding contributions from $z < 0.05$ appears well justified.

\subsection{Proton-lead collisions}
As of now, $W^\mp D^{(*)\pm}$ production has not yet been measured in $p$Pb collisions. In this Section we discuss predictions for cross sections and $R_c^\pm$  calculated using our framework and, more importantly, the expected statistical significance 
with a realistically achievable luminosity.

The efficiency $\epsilon$ of particle detectors is the ratio of the number of detected events to the true number of events:
\begin{equation}
    \label{eq: N_expected}
    N_\text{detected} = \epsilon N_\text{true}.
\end{equation}
For a given detector, this quantity is not process independent. However, a reasonable expectation is that it should not change too much by changing the colliding particles. The true number of events can be expressed using the integrated luminosity of the experiment $L_\text{int}$ and the measured cross section $\sigma$:
\begin{equation}
    \label{eq: N_true}
    N_\text{true} = L_\text{int} \sigma.
\end{equation}
The statistical error of the measurement is given by
\begin{equation}
    \label{eq: delta sigma}
    \delta \sigma = \frac{\sigma}{\sqrt{N_\text{detected}}}.
\end{equation}
Combining equations \eqref{eq: N_expected}, \eqref{eq: N_true} and \eqref{eq: delta sigma}, we can write the efficiency of the ATLAS detector for measuring $W^\mp D^{(*) \pm}$ events as
\begin{equation*}
    \epsilon_\text{ATLAS} (W^\mp D^{(*)\pm}) = \frac{\sigma_{pp} (W^\mp D^{(*)\pm})}{[\delta \sigma_{pp} (W^\mp D^{(*)\pm})]^2 L_\text{int} (pp)}
    \approx
    \begin{cases}
        0.002 \quad (D^\pm)\\
        0.009 \quad (D^{*\pm})
    \end{cases}
    .
\end{equation*}
The quantities on the right-hand side of the equation correspond to the integrated ATLAS $pp$ data \cite{measurementforcomparison}. Using this form for the efficiency of the detector, we get an estimate for the expected statistical uncertainty in a $p$Pb experiment:
\begin{equation*}
    \delta \sigma_{p\text{Pb}} (W^\mp D^{(*) \pm}) = \sqrt{\frac{\sigma_{p\text{Pb}} (W^\mp D^{(*)\pm})}{\epsilon_\text{ATLAS} (W^\mp D^{(*)\pm}) L_\text{int} (p\text{Pb})}}.
\end{equation*}
For the cross sections $\sigma_{p\text{Pb}} (W^\mp D^{(*)\pm})$ we use theoretical predictions evaluated with the EPPS21NLO \cite{EPPS21} and nNNPDF3.0NLO \cite{AbdulKhalek:2022fyi} nuclear PDFs, as we are confident the predictions will estimate the experimental value well enough. We also re-evaluate the $pp$ cross sections with the proton-baseline PDFs of these nuclear PDFs (CT18A for EPPS21). As the LHAPDF library gives per-nucleon PDFs, the cross sections obtained from MCFM are still multiplied by 208 to find the cross sections at the ${p\text{Pb}}$ level. We take as the integrated luminosity of the $p$Pb experiment $L_\text{int} (p\text{Pb}) = 1.2\text{pb}^{-1}$, which has been suggested as a reasonable goal for ATLAS \cite{citron2019futurephysicsopportunitieshighdensity}. We will proceed by choosing $\sqrt{s} = 8.5$ TeV and considering the same kinematic cuts as was used for $pp$ collisions, outlined in Table \ref{tab: cuts}.

\begin{figure}[b!]
    \centering
    \includegraphics[width=0.49\linewidth]{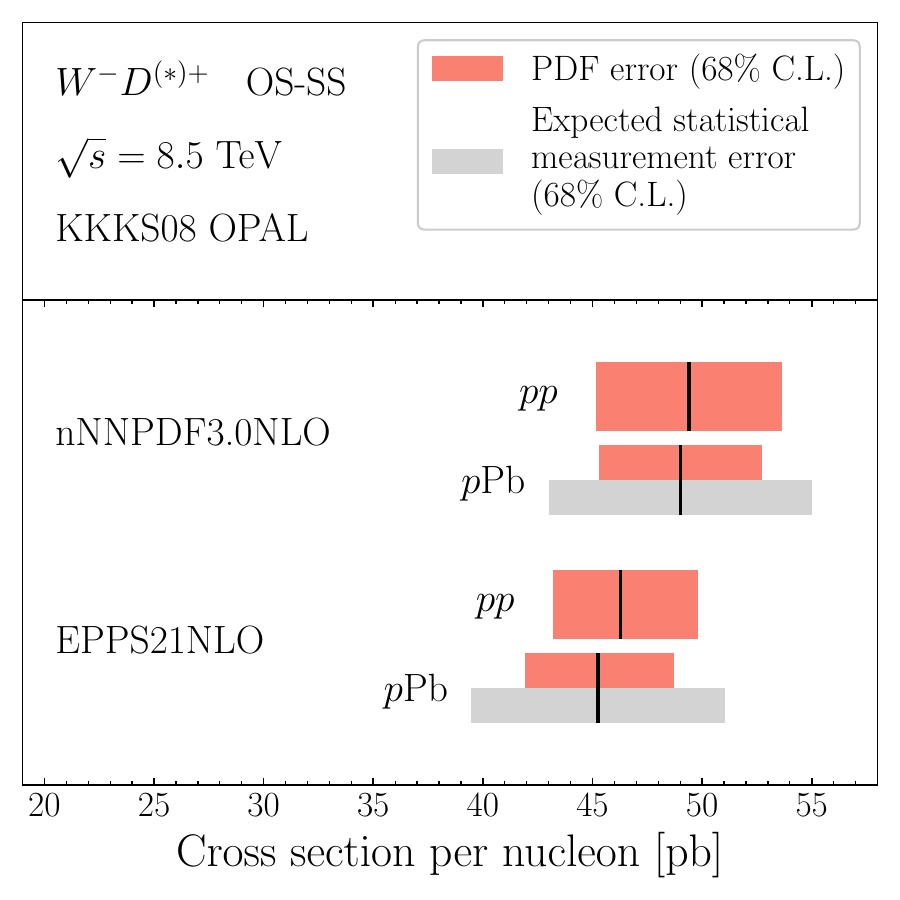}
    \includegraphics[width=0.49\linewidth]{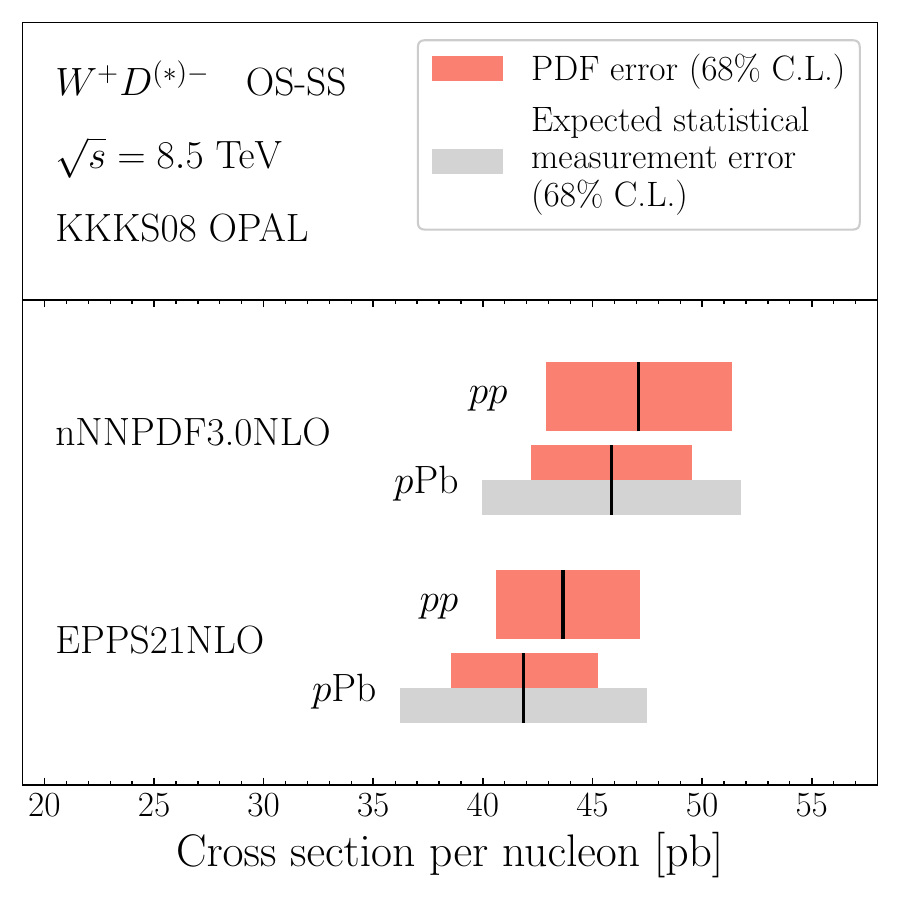}
    \caption{Integrated cross sections $\sigma(W^\mp D^\pm) + \sigma(W^\mp D^{*\pm})$ in $pp$ and $p$Pb collisions. }
    \label{fig: pPb cross sections}
\end{figure}

Figure \ref{fig: pPb cross sections} shows values for integrated cross sections in $pp$ and $p$Pb collisions, where final states involving $D^\pm$ and its excited state $D^{*\pm}$ are summed over. This is done to increase the statistics of the projected experiment. For $p$Pb collisions, the expected statistical measurement uncertainty is shown in gray. This corresponds to roughly 100 reconstructed events. We see that this uncertainty is of the same order of magnitude as the 68\% C.L. PDF uncertainty. 

\begin{figure}[htb!]
    \centering
    \includegraphics[width=0.5\linewidth]{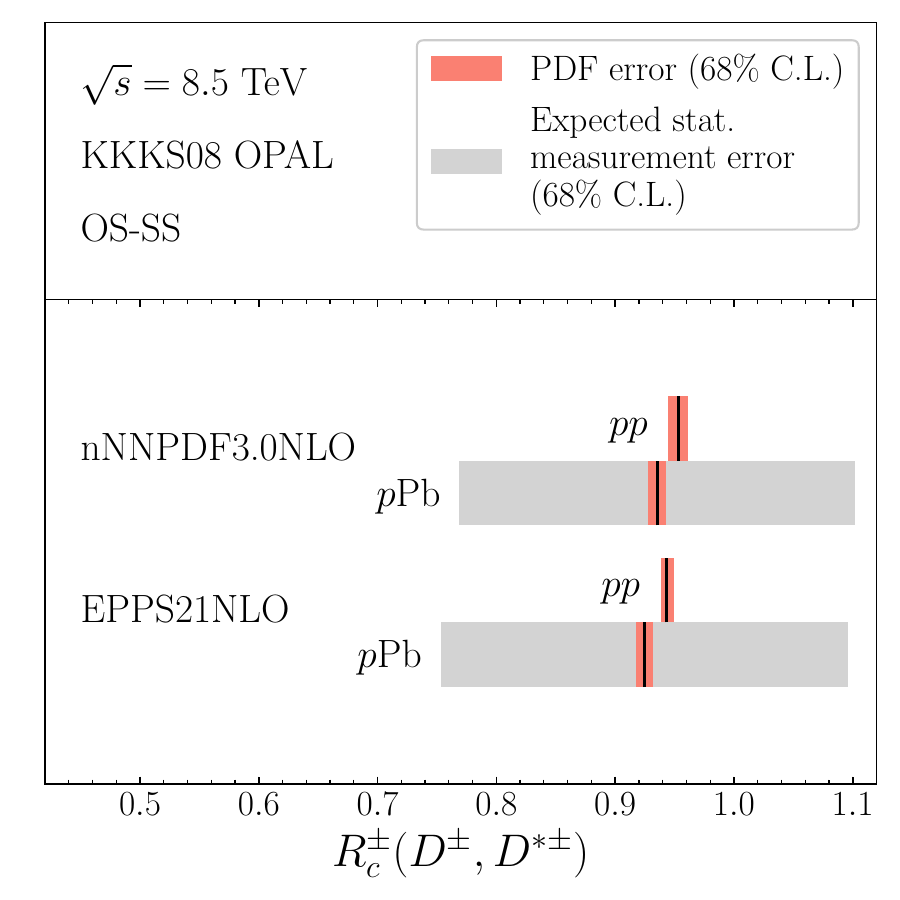}
    \caption{The production ratio $R_c^\pm (D^\pm, D^{*\pm})$ in $pp$ and $p$Pb collisions.}
    \label{fig: pPb Rcpm}
\end{figure}

Figure \ref{fig: pPb Rcpm} shows values for the production ratio $R_c^\pm$ in $pp$ and $p$Pb collisions. The predicted values are lower in $p$Pb in comparison to $pp$ collisions, which can be attributed to the fact that the per-nucleon valence $d$ PDF of the lead nucleus is larger than the corresponding proton PDF due to the presence of neutrons. The estimated statistical uncertainty is, however, way larger than the current PDF uncertainty. We note that the smallness of the PDF uncertainty here is partly related to the assumption $s_- = 0$ made in EPPS21 and nNNPDF3.0 fits.

\section{Conclusions}

We have presented a comprehensive analysis of associated $W^\mp D^{(*)\pm}$ production in $pp$ and $p$Pb collisions at NLO within the collinearly factorized perturbative QCD and GM-VFNS. We explained in detail how results for $W + c$ production from parton-level programs such as MCFM or MadGraph can be convoluted with scale dependent FFs to compute $W^\mp D^{(*)\pm}$ cross sections consistently subtracting the logarithmic contributions originating from collinear final-state radiation which are resummed to the scale dependence of FFs. This process retains the full dependence of the charm-quark mass which should be relevant at low-enough values of transverse momenta. The earlier calculations within collinear factorization and FFs, extending to NNLO in perturbative QCD, have been performed by neglecting the charm quark mass from the beginning and our results can also be seen as a reliability check of that approach at the considered kinematics.

Our numerical studies concentrated mostly on the $\sqrt{s} = 13\,\text{TeV}$ $pp$ collisions and the kinematic cuts were matched to the recent measurement from the ATLAS collaboration \cite{measurementforcomparison} which reconstructed $D$ mesons above transverse momenta of $8\,{\rm GeV}$. We noticed that the absolute integrated and differential NLO cross sections underestimate the measurements by some 20\% when using FFs fitted to high-energy LEP data. This is consistent with the zero-mass results i.e. the charm-mass effects do not seem to play a major role at these values of transverse momenta. This gives a reason to assume that the zero-mass NNLO calculations should also be relatively inert to residual mass effects and that they can therefore be reliably used in PDF fits at the NNLO level. In passing, we noticed that the ratios between $W^\mp D^{\pm}$ and $W^\mp D^{*\pm}$ cross sections are generally lower than the experimental values which, in turn, are consistent with the charm-hadron production fractions measured at LEP. This indicates that these ratios -- as a function of $p_{T, D}$ which also gives a handle on the $z$ dependence -- could be useful for future fits of $D$-meson FFs. In addition, the $W^\pm + {\rm charm}$ processes should be useful in further understanding the recently noticed differences in the effective $c \rightarrow {\rm charmed\text{-}hadron}$ branching fractions between $e^+e^-$ and $pp$ collisions. To this end, we propose to measure the $W^\pm + {\rm charmed \ hadron}$ processes down to zero transverse momentum, if experimentally feasible. 

In order to isolate the potential constraints on PDFs we considered production ratios of events involving $W^+$ to events involving $W^-$. These were quantitatively shown to eliminate most of the theory uncertainties like those originating from scale variations, FFs, ambiguities associated with the fragmentation variable and intrinsic charm. These ratios also cancel several systematic experimental uncertainties. Out of the three proton PDFs considered, CT18A and MSHT20 provided a reasonable correspondence with the recent ATLAS data, while NNPDF4.0 underestimated the production ratios in most of the considered kinematic regimes. The main reason was traced to the strangeness asymmetry $s_- = s - \overline{s}$ which is zero in CT18A but which is fitted in MSHT20 and NNPDF4.0. This was further studied through PDF reweighting methods which confirmed that an optimal agreement with the ATLAS data would require a significantly smaller $s_-$ in NNPDF4.0. Alternatively, an increased total strange PDF $s_+ = s + \overline{s}$ would also improve the situation but out of these two $s_-$ is not so well constrained by other data. However, the required changes in NNPDF4.0 would be rather significant which could potentially lead to conflicts with other data included in these global fits. Whether there is a compromise remains to be seen. For example, the dimuon production in neutrino-nucleus DIS has been a traditional cornucopia of strange-quark PDFs and the recent theoretical improvements on this side \cite{Helenius:2025fpy,Helenius:2026uuz,Caola:2026kvl} could also affect the situation. 

Finally, we also considered the prospects of measuring $W^\mp D^{(*)\pm}$ cross sections in $p\text{Pb}$ collisions at the LHC. We found that an integrated luminosity of $1.2\text{pb}^{-1}$ at $\sqrt{s} = 8.5~\text{TeV}$ with realistic efficiency estimate would lead to around 100 events of both charge combinations. This corresponds to a statistical error which is of the same order as the $68 \%$ C.L. PDF uncertainty from EPPS21 and nNNPDF4.0. However, for the ratios of events involving $W^+$ to events involving $W^-$ the expected statistical uncertainty is way larger than the PDF uncertainties due to cancellation of PDF uncertainties upon taking the ratio. In this respect, a more detailed study of the impact of possible non-zero strangeness asymmetry in nuclear PDFs would still be required.

\acknowledgments

The financial support from the Research Council of Finland Project No.~361179 (I.H.), and the Center of Excellence in Quark Matter of the Research Council of Finland, project 346326, are acknowledged. This work has utilized the computing infrastructure of the Finnish IT Center for Science (CSC), under the project jyy2580. 


\bibliographystyle{JHEP}
\bibliography{biblio.bib}

\end{document}